\documentclass[11pt,a4paper]{article}
\usepackage{jcappub}
\usepackage{multirow}

\title{Spatially Homogeneous Einstein-Aether Cosmological Models: Scalar Fields with a Generalized Harmonic Potential}

\author[a]{B. Alhulaimi,}
\author[b]{R. J. van den Hoogen,}
\author[a]{A. A. Coley}

\affiliation[a]{Department of Mathematics, \\
Dalhousie University,\\
6316 Coburg Road, Halifax, N.S., B3H 4R2, Canada}
\affiliation[b]{Department of Mathematics, Statistics and Computer Science, \\
St. Francis Xavier University,\\
2323 Notre Dame Avenue, Antigonish, N.S., Canada}

\emailAdd{bs748397@dal.ca}
\emailAdd{rvandenh@stfx.ca}
\emailAdd{aac@mathstat.dal.ca}

\abstract{
Inflationary spatially homogeneous cosmological models within an Einstein-Aether gravitational framework are investigated.  The matter source is assumed to be a scalar field which is coupled to the aether field expansion and shear scalars through the generalized harmonic scalar field potential.   The evolution equations are expressed in terms of expansion-normalized variables to produce an autonomous system of ordinary differential equations suitable for numerical and local stability analysis.  An analysis of the local stability of the equilibrium points indicate that there exists a range of values of the parameters in which there exists an accelerating expansionary future attractor.}

\arxivnumber{1234.5678}

\begin{document}
\maketitle


\section{Introduction} \label{introduction}

\subsection{Lorentz-violating Inflationary Cosmological Models}

Lorentz invariance is a fundamental symmetry of General Relativity (GR) and standard particle physics and has been tested to a very high degree of accuracy.  However, there is growing evidence that some issues in quantum gravity (certain divergences, micro-causality) can be resolved by the removal of Lorentz invariance \cite{Mattingly:2005re,Jacobson:2000xp}. Furthermore, some approaches to quantum gravity may even desire a preferred rest frame in vacuum \cite{Liberati:2013xla,Jacobson:2008aj}.  On larger scales, explanations of Dark Energy and Dark Matter in the current cosmological paradigms based on GR might also be explained using an alternative theory of gravity \cite{Zlosnik:2006zu,Clifton:2011jh}, such as those in which the Lorentz invariance requirement is relaxed. Indeed, in cosmology there is a natural frame associated with the cosmic microwave background, and therefore it is possible the Lorentz invariance assumption may be relaxed at late times.

The Einstein-Aether (AE) theory \cite{Jacobson:2000xp,Jacobson:2008aj,Jacobson:2004ts,Jacobson:2010mxa,Jacobson:2010mxb,Garfinkle:2007bk,Garfinkle:2011iw} is a proposed model of gravity which preserves general covariance and incorporates a violation of Lorentz invariance. The local space-time structure is determined by a dynamical time-like vector field, $u^a$ (the aether), together with a metric tensor, $g_{ab}$. The field equations for this Einstein-Aether theory essentially consist of GR with a modified source due to the aether together with additional field equations describing the evolution of the aether vector field.  In standard Einstein-Aether theory, it is commonly assumed that the violation of Lorentz invariance is only within the gravity sector of the theory, while the matter sector continues to be coupled only to the metric, and hence remains Lorentz invariant.  However, it is natural to expect that Lorentz violations in the matter sector could also be permitted, albeit with the understanding that there are quite stringent constraints on such Lorentz violations \cite{Mattingly:2005re}.

Assuming matter is determined by a scalar field, some researchers have investigated the potential changes that arise as a result of a violation of Lorentz invariance in the matter sector of the Einstein-Aether theory \cite{Barrow:2012qy,Sandin:2012gq,Donnelly:2010cr,Solomon:2013iza,Solomon:2015hja,Alhulaimi:2017,Kanno:2006ty}. Barrow \cite{Barrow:2012qy} investigated the effect of scalar field/aether field coupling in which the dependence of the scalar field in the potential was exponential in nature.  He found there are solutions with the possibility that the coupling parameter enables inflation in situations in which it would not otherwise occur. Sandin {\em et al.} used a similar ansatz for the scalar field potential in \cite{Barrow:2012qy} and determined that there is a fundamental change in the future asymptotic behaviour when the coupling parameter becomes sufficiently large.  Donnelly and Jacobson \cite{Donnelly:2010cr} considered a chaotic inflationary scenario and determined that the coupling of the scalar field to the aether field can either slow down or speed up the evolution of inflation. Solomon and Barrow \cite{Solomon:2013iza} completed a detailed analysis with no prescribed coupling between the scalar field and the aether field and found conditions on the potential that must be satisfied if one is to have stable slow-roll inflationary solutions.  Alhulaimi \cite{Alhulaimi:2017} generalized some of these results to include not only a coupling of the scalar field to the aether vector expansion, but also a coupling to the aether vector shear.  Where most others have coupled the scalar field to the aether field through the scalar field potential, Kanno and Soda \cite{Kanno:2006ty} took a very different approach.   In their analysis, they assumed that the aether parameters $c_i$ in equation \eqref{K} are functions of the scalar field.  They found that it is possible to have inflation without a scalar field potential,  i.e., with a massless scalar field.

Inflation has become a well accepted, but not yet proven, mechanism which attempts to explain many cosmological issues \cite{Olive:1989nu,Linde:1987}. A finite period of accelerated expansion (inflation) in the early universe is desirable to help address the isotropy, spatial homogeneity, horizon, and flatness problems \cite{Olive:1989nu,Linde:1987}.   The standard inflationary model consists of a single massive scalar field that causes the universe to experience a period of exponential expansion early in its evolution.  A common inflationary scenario assumes a convex potential, such as a harmonic scalar field potential, in which inflation takes place during a period of ``slow roll'', when the scalar field is decreasing very slowly in comparison to the expansion of the universe.  Not only is a period of accelerated expansion desirable at early times, but due to the Dark Energy phenomenon\cite{Riess:1998cb,Perlmutter:1998np}, a period of accelerated expansion is also an attractive property to have at late times.

In this paper we shall investigate the dynamical evolution of a class of isotropic and anisotropic spatially homogeneous Einstein-Aether cosmological models containing a scalar field that is coupled to the aether field through the scalar field potential.  In particular, we explore the potential impact of Lorentz violation in the matter sector on the standard inflationary scenario \cite{Olive:1989nu,Linde:1987}. More precisely, we study the inflationary scenario and investigate whether the inflationary solutions proposed \cite{Donnelly:2010cr,Alhulaimi:2017} are stable when spatial curvature and anisotropy perturbations are considered. Further, we are also interested in the possibility of late time accelerated expansion in these models.

\subsection{Einstein-Aether Gravity}

The action under consideration contains a Lagrangian describing Einstein-Aether gravity together with a Lagrangian for a matter field or fields (\textsc{M})
\begin{equation}
S=\int d^{4}x\sqrt{-g}\left[\frac{1}{8\pi G} {\mathcal L}^{\textsc{AE}}+{\mathcal L}^{\textsc{M}} \right] .\label{action}
\end{equation}
The lagrangian ${\mathcal L}^{\textsc{AE}}$ depends on the spacetime metric, $g_{ab}$, and the normalized aether vector field, $u^a$, and has the form  \cite{Jacobson:2010mxa,Jacobson:2000xp,Jacobson:2010mxb,Garfinkle:2011iw,Garfinkle:2007bk,Jacobson:2008aj,Jacobson:2004ts,Donnelly:2010cr,Solomon:2013iza,Coley:2015qqa}:
\begin{equation}
{\mathcal L}^{\textsc{AE}}=\frac{1}{2}R - K^{ab}_{\phantom{ab} cd}\nabla_{a}u^c\nabla_{b}u^d  + \lambda(u^au_a+1) \label{Lagrangian_AE}
\end{equation}
where
\begin{equation}
K^{ab}_{\phantom{ab}{cd}}  \equiv  c_1 g^{ab} g_{cd} + c_2\delta_{c}^{a} \delta_{d}^{b}+c_3\delta_{d}^{a}\delta_{c}^{b}+ c_4 u^{a} u^{b} g_{cd}.\label{K}
\end{equation}
We note that the parameters $c_i$ defined here are the same as those used in \cite{Coley:2015qqa} which are equal to half of the values of the $c_i$ employed in \cite{Garfinkle:2011iw,Garfinkle:2007bk} with an opposite sign for the $c_4$. In comparison to \cite{Solomon:2013iza}, the $c_i$ used here are $8\pi G$ times the values of $c_i$. The metric signature is assumed to be $+2$.

Let
\begin{equation}
{\mathcal L}^{\textsc{U}}= - K^{ab}_{\phantom{ab} cd}\nabla_{a}u^c\nabla_{b}u^d
\end{equation}
then the variation of the action \eqref{action} with respect to the aether vector field $u^a$ yields
\begin{equation}
-2\lambda u_a =  \frac{\delta {\mathcal L}^{\textsc{U}}}{\delta u^a}+8\pi G\,\frac{\delta {\mathcal L}^{\textsc{M}}}{\delta u^a},\label{deltaL-deltau}
\end{equation}
which when contracted with $u^a$, provides an explicit expression for the Lagrange multiplier
\begin{equation}
2\lambda  =  \frac{\delta {\mathcal L}^{\textsc{U}}}{\delta u^a}u^a+8\pi G\,\frac{\delta {\mathcal L}^{\textsc{M}}}{\delta u^a}u^a. \label{lambda}\\
\end{equation}
Equation \eqref{lambda} can then be used to eliminate the contribution of the Lagrange multiplier $\lambda$ when calculating the effective energy momentum tensors.

Variation of the action \eqref{action} with respect to $g^{ab}$, $\lambda$ and a generalized matter field or fields $\Psi$, yields
\begin{eqnarray}
G_{ab}   &=& T_{ab}^{\textsc{U}} + 8\pi G \,T_{ab}^{\textsc{M}},\\
u^a u_a &=& -1, \\
\frac{\delta {\mathcal L}^{\textsc{M}}}{\delta \Psi} &=& 0.\label{deltaL-deltapsi}
\end{eqnarray}
When the contributions from the Lagrange multiplier in equation \eqref{lambda} are taken into account, expressions for the effective energy momentum tensors due to the aether vector field and the matter field become
\begin{eqnarray}
T_{ab}^{\textsc{U}}  &=& -2\frac{\delta{\mathcal L}^{\textsc{U}}}{\delta g^{ab}} + g_{ab}{\mathcal L}^{\textsc{U}}+\frac{\delta{\mathcal L}^{\textsc{U}}}{\delta u^{c}}u^cu_au_b, \label{Matter-AE}\\
T_{ab}^{\textsc{M }} &=& -2\frac{\delta{\mathcal L}^{\textsc{M}}}{\delta g^{ab}} + g_{ab}{\mathcal L}^{\textsc{M}}+\frac{\delta{\mathcal L}^{\textsc{M}}}{\delta u^{c}}u^cu_au_b. \label{Matter-M}
\end{eqnarray}

Given the form of the Lagrangian in equation \eqref{Lagrangian_AE}, the effective energy momentum tensor due to the Aether field \cite{Jacobson:2004ts,Garfinkle:2011iw,Garfinkle:2007bk,Donnelly:2010cr,Solomon:2013iza,Coley:2015qqa} is
\begin{eqnarray}
T_{ab}^{\textsc{U}} &=& 2\nabla_{c}\Bigl(J_{(a}^{\phantom{a}c}u^{\phantom{c}}_{b)} - J^c_{\phantom{c} (a}u^{\phantom{c}}_{b)} -J_{(ab)}u^c \Bigr) \nonumber\\
            &&  2c_1\Bigl((\nabla_{a} u^c)(\nabla_{b} u_c) - (\nabla^c u_a)(\nabla_c u_b)\Bigr) -2c_4\dot{u}_a\dot{u}_b \nonumber\\
            && -2\left( u^d\nabla_c J^{c}_{\phantom{c}d}+c_4\dot{u}_c\dot{u}^c   \right)u_au_b - g_{ab} \Bigl(K^{cd}_{\phantom{cd}ef} \nabla_{c} u^e \nabla_{d} u^f\Bigr),\label{T_ab_AE}
\end{eqnarray}
where
\begin{eqnarray}
J^{a}_{\phantom{a}b}  &=& -K^{ac}_{\phantom{ac}bd} \nabla_{c} u^d,\\
\dot{u}^a &=& u^b\nabla_b u^a.
\end{eqnarray}

\subsection{Matter as a Scalar Field}

Assuming that the matter component of the universe is a single scalar field having a potential that is assumed to be a function of the scalar field together with the expansion and shear scalars of the aether vector field, the matter Lagrangian becomes:
\begin{equation}
L^\textsc{{M}}= -\frac{1}{2} g^{a b } \nabla_{a}\phi \nabla_{b}\phi - V(\phi,\theta,\sigma^2),
\end{equation}
where $\theta=\nabla_a u^a$ is the expansion scalar and $\sigma^2=\frac{1}{2}\sigma_{ab}\sigma^{ab}$ is the shear scalar.  Again taking into account contributions from the Lagrange multiplier, equation \eqref{Matter-M} yields the effective energy momentum tensor due to the scalar field
\begin{eqnarray}
T_{ab}^{\textsc{M}} &=&  \nabla_a \phi \nabla_b \phi - \left(\frac{1}{2} \nabla_a\phi \nabla^a \phi + V \right)g_{ab}
+ \theta V_\theta g_{ab}+\dot{V}_\theta h_{ab} \nonumber\\
&& +\left(\theta V_{\sigma^2}+\dot{V}_{\sigma^2}\right)\sigma_{ab}+V_{\sigma^2}\dot{\sigma}_{ab}-2\sigma^2V_{\sigma^2}u_au_b
\label{scalar_T}
\end{eqnarray}
where the terms $V_\theta$ and $V_{\sigma^2}$ are the partial derivatives of the scalar field potential with respect to $\theta$ and $\sigma^2$, respectively, and $h_{ab}\equiv g_{ab}+u_au_b$.  If there is no coupling between the aether field and the scalar field via the potential, then $V_\theta=V_{\sigma^2}=0$ and the energy momentum tensor reduces to the standard form for a minimally coupled scalar field.
In addition, the field equation \eqref{deltaL-deltapsi} yields the Klein-Gordon equation for the scalar field:
\begin{equation}
\nabla^a\nabla_a\phi-V_{\phi}=0.\label{GeneralKleinGordoneqaution}
\end{equation}


\section{Isotropic Einstein-Aether Models coupled to a Scalar Field} \label{section2}

\subsection{The Spatially Homogeneous and Isotropic Model}

We shall assume that the spacetime is spatially homogeneous and isotropic having spacetime coordinates $[t,r,\theta,\psi]$ and a metric of the form:
\begin{equation}
ds^2=-dt^2+a(t)^2\left(\frac{1}{1-kr^2}dr^2+r^2d\theta^2+r^2\sin^2(\theta)d\psi^2\right),
\end{equation}
where $k$ takes on values $\{-1,0,1\}$ for negative, zero and positive spatial curvature, respectively.  In a spatially homogeneous and isotropic cosmological model with comoving time, the aether vector field necessarily coincides with the rest frame defined by the Hubble expansion.  Specifically, this implies that in spatially homogeneous and isotropic models that the aether vector must be orthogonal to the three-dimensional spatial hyper-surfaces and takes the form $u^a=(1,0,0,0)$.

With the above assumptions on the metric and the aether vector, the shear, the vorticity and the acceleration of the aether vector are zero and the covariant derivative
\begin{equation}
\nabla_b u_{a}=\frac{1}{3}\theta (g_{ab}+u_au_b)
\end{equation}
is simply determined by the expansion scalar
\begin{equation}
\theta = \nabla_au^a = 3\frac{\dot a}{a}.
\end{equation}

With the definition of $T_{ab}^{\textsc{U}}$ in equation \eqref{T_ab_AE}, the effective energy density $\rho^{\textsc{U}}$ and isotropic pressure, $p^{\textsc{U}}$,   due to the aether field are
\begin{eqnarray}
\rho^{\textsc{U}} & = & -\frac{1}{3}c_\theta\theta^2 ,\\
p^{\textsc{U}} & = & \frac{1}{3}c_\theta\theta^2 +\frac{2}{3}c_\theta\dot\theta.
\end{eqnarray}
Where a new parameter $c_{\theta}= (c_1 +3c_2 +c_3)$, defined before in \cite{Coley:2015qqa,Latta:2016jix}, allows for some efficiencies in notation since the field equations are independent of any other linear combinations of the $c_i$.
The Einstein-aether field equations reduce to the following:
\begin{eqnarray}
0&=& -\frac{1}{3}(1+c_\theta)\theta^2+8\pi G\rho^{\textsc{M}}-\frac{3k}{a^2}, \label{Friedmann}\\
0 &=& -(1+c_\theta)\dot\theta -\frac{1}{3}(1+c_\theta) \theta^2  -\frac{8\pi G}{2}(\rho^{\textsc{M}}+3p^{\textsc{M}} ),\label{Raychaudhuri}
\end{eqnarray}
where there still exists the freedom to choose some appropriate units.   Without loss of generality, new units can be chosen so that $\frac{8\pi G}{1+c_\theta}=1$ in which case the explicit dependence of the field equations on the aether parameter $c_\theta$ has been eliminated.

\subsection{The Scalar Field Potential}
We shall consider a class of quadratic scalar field  potentials  of the form
\begin{equation}
V(\phi, \theta)=\frac{1}{2} m^2 \phi^2 + \mu \theta \phi,
\end{equation}
where the scalar field/Aether field coupling term $\mu\theta\phi$ term can be interpreted as an incorporation of an external source, in this case the Aether, acting on the scalar field.  The effective energy density $\rho^{\textsc{M}}$ and isotropic pressure $p^{\textsc{M}}$ from equation \eqref{scalar_T}, are
\begin{eqnarray}
\rho^{\textsc{M}} & = & \frac{1}{2}\dot\phi^2+ \frac{1}{2} m^2 \phi^2,\\
p^{\textsc{M}} & = & \frac{1}{2}\dot\phi^2- \frac{1}{2} m^2 \phi^2+\mu\dot\phi.
\end{eqnarray}
The Klein-Gordon equation becomes
\begin{equation}
0=\ddot{\phi}+\theta\dot{\phi}+m^2\phi+\mu\theta,\label{KG}
\end{equation}
where we can more clearly see how $\mu\theta$ acts like an external source in the Klein-Gordon equation when compared to the usual non-coupled $\mu=0$ version of the equation.

\subsection{The Dynamical System}

The Einstein-aether field equations and the Klein-Gordon equation can be expressed as the following system of ordinary differential equations
\begin{eqnarray}
\dot\theta &=& -\frac{1}{3} \theta^2 +\frac{m^2}{2}\phi^2-\psi^2-\frac{3}{2}\mu\psi,\label{DS1}\\
\dot\phi&=&\psi,\\
\dot\psi&=& -\theta\psi-m^2\phi-\mu\theta.\label{DS3}
\end{eqnarray}
with first integral
\begin{equation}
\frac{\theta^2}{3} = \frac{m^2}{2}\phi^2+\frac{1}{2}\psi^2-\frac{3k}{a^2}. \label{Friedmann2}\\
\end{equation}

Equations \eqref{DS1}-\eqref{DS3}, therefore, yield a three dimensional dynamical system for the variables $(\theta, \phi,\psi)$  depending on three parameters $(k, m, \mu)$ having a first integral given by equation \eqref{Friedmann2}. Since the system of equations is invariant under the transformation $(\mu,\phi,\psi) \mapsto -(\mu,\phi,\psi)$, we can without loss of generality, assume that $\mu\geq 0$.  Given that the phase space for the dynamical system defined in equations \eqref{DS1}-\eqref{DS3} with first integral \eqref{Friedmann2} is not  bounded, we employ dimensionless variables \cite{Coley:2003mj,wainwright_ellis2005} which will transform the system into an autonomous system of differential equations on a bounded phase space.

\subsection{Qualitative Analysis}

\subsubsection{Introducing Normalized Variables}

Introducing a time variable  $\tau$
\begin{equation}
\frac{d\tau }{dt}=\sqrt{1+\theta^2}
\end{equation}
and normalized variables
\begin{eqnarray}
    D &\equiv& \frac{\theta}{\sqrt{1+\theta^2}}, \\
 \Phi &\equiv& \sqrt{\frac{3}{2}}\left(\frac{m \phi}{\sqrt{1+\theta^2}}\right),\\
 \Psi &\equiv& \sqrt{\frac{3}{2}}\left(\frac{\dot{\phi}}{\sqrt{1+\theta^2}}\right),
 \label{1BD}
\end{eqnarray}
the evolution equations \eqref{DS1}-\eqref{DS3} become
\begin{eqnarray}
   D^{\prime} &=& (1-D^2)\mathcal{X},  \label{3De1}\\
\Phi^{\prime} &=& m \Psi \sqrt{1-D^2}-D \Phi \mathcal{X},  \label{3De2}\\
\Psi^{\prime} &=& -D \Psi- \sqrt{1-D^2} \left( m \Phi+\sqrt{\frac{3}{2}}\mu D \right) -\Psi D \mathcal{X},\label{3De3}
\end{eqnarray}
where the prime here indicates the differentiation with respect to $\tau$ and $\mathcal{X}$ is given by the expression
\begin{eqnarray}
\mathcal{X} &=& \frac{\dot{\theta}}{\theta^2+1}\nonumber\\
&=& -\frac{1}{3}D^2-\frac{2}{3}\Psi^2+\frac{1}{3} \Phi^2 -\sqrt{\frac{3}{2}}\mu \Psi \sqrt{1-D^{2}}
\label{X2}
\end{eqnarray}

The Friedmann equation \eqref{Friedmann2} becomes
\begin{equation}
D^2-\Phi^2-\Psi^2= -\frac{9 k}{ a^2(1+\theta^2)}.
\label{FE1}
\end{equation}
Further, if $k = -1,0$, then it follows that
\begin{equation}
0 \leq \Phi^2+\Psi^2  \leq D^2\leq 1.\label{Friedmann_Constraintfrw}
\end{equation}
That is, $D, \Phi, \Psi $ are bounded in the flat and negatively curved scenarios and the phase space is a compact set.  Hence forward, we shall restrict our analysis to $k=0, -1$ cases only.

\subsubsection{Invariant Sets and Monotonic Functions}

The phase space can be subdivided into four disjoint invariant sets according to the curvature of the model and whether $D=1$ ($\theta\to\infty$) or not.  A superscript ``$-$'' indicates that points in this set represent negatively curved models, while a superscript ``$0$'' indicates a flat model. The invariant sets are
\begin{eqnarray*}
\mbox{FRW}^- &=& \{(D,\Phi,\Psi)|D<1,\Phi^2+\Psi^2  <   D^2\},\\
\mbox{FRW}^0 &=& \{(D,\Phi,\Psi)|D<1,\Phi^2+\Psi^2  =   D^2\},\\
\mbox{D}^- &=& \{(D,\Phi,\Psi)|D=1,\Phi^2+\Psi^2  < 1\},\\
\mbox{D}^0 &=& \{(D,\Phi,\Psi)|D=1,\Phi^2+\Psi^2  = 1\},
\end{eqnarray*}
the dimensions of which are 3, 2, 2, and 1, respectively.  We note the following closure properties of the sets
\begin{eqnarray*}
\overline{\mbox{FRW}^-}&=&\mbox{FRW}^-\cup\mbox{FRW}^0\cup\mbox{D}^-\cup\mbox{D}^0,\\
\overline{\mbox{D}^-} &=&\mbox{D}^-\cup\mbox{D}^0,\\
\overline{\mbox{FRW}^0}&=&\mbox{FRW}^0\cup\mbox{D}^0.
\end{eqnarray*}
Further the invariant set $\mbox{D}^-$ can be divided into three distinct pieces depending on whether $\Phi<0$, $\Phi=0$ or $\Phi>0$.

If we define $\Lambda_1 = D^2- \Phi^2-\Psi^2$ and $\Lambda_2=D^2-1$ then
\begin{eqnarray}
 \frac{\Lambda_1^{\prime}}{\Lambda_1} &=& -\frac{2}{3}  D  ( 3\mathcal{X}+1 ),\\
 \frac{\Lambda_2^{\prime}}{\Lambda_2} &=& 2D\mathcal{X}.
\end{eqnarray}
in which case the non-negative function $W=(\Lambda_1)^2(\Lambda_2)^2$ has the derivative
\begin{equation}
W^\prime = -\frac{4}{3}WD.
\end{equation}
Since $W> 0$ and $W^\prime<0$  in the set $\mbox{FRW}^-$ we can conclude that there are no periodic orbits in this 3-dimensional invariant set. This implies that there are no equilibrium points in the set $\mbox{FRW}^-$, and any equilibrium points of the autonomous system of differential equations \eqref{3De1}-\eqref{3De3} will lie in the lower dimensional invariant sets $\mbox{FRW}^0$, $\mbox{D}^-$  or $\mbox{D}^0$.  We also note that in the invariant set $\mbox{D}^-\cap\{\Phi<0\}$, one can show that $\Phi^\prime <0$, and similarly in the set $\mbox{D}^-\cap\{\Phi>0\}$, one can show that $\Phi^\prime >0$.  This shows that there are no closed or periodic orbits in these sets.  The remaining portion $\mbox{D}^-\cap\{\Phi=0\}$, is 1 dimensional. No monotonic function has been found in the set $\mbox{FRW}^0$ and consequently the most interesting qualitative behaviour for this autonomous system of differential equations occurs in $\overline{\mbox{FRW}^0}$.

\subsubsection{Equilibrium Points}
The equilibrium points and a non-isolated line of equilibria for the system \eqref{3De1}-\eqref{3De3} with the value of $\mathcal{X}$  and their stability are summarized in Table \eqref{Table1}.
\begin{table}
\begin{center}
\begin{tabular}{|c|c|c|l|l|l|c|c|}
\hline
Pt      & $(D,\Phi,\Psi)$  & $\mathcal{X}$   &\multicolumn{3}{|c|}{Stability} & Invariant    & $q$ \\ \cline{4-6}
 	
 	    &                  &                 & $\mu < {\mu_{c}}$ & $\mu= {\mu_{c}}$ & $\mu> {\mu_{c}}$ &  Set  &     \\
\hline
$P_{0}$ &  $(0,0, 0)$      & $0$              &	 Sink    & Sink    & Saddle   & $\mbox{FRW}^0$  & DNE \\
$P_{1}$ &  $(1, 0,1)$      & $-1$             &  Source	 & Source  & Source   & $\mbox{D}^0$    & $q>0$ \\
$P_{2}$ &  $(1,0,-1)$      & $-1$             &  Source	 & Source  & Source   & $\mbox{D}^0$    & $q>0$ \\
$P_{3}$ &  $(1,1, 0)$      & $0$              &  Saddle	 & Saddle  & Saddle   & $\mbox{D}^0$    & $q<0$ \\
$P_{4}$ &  $(1,-1,0)$      & $0$              &  Saddle	 & Sink    & Sink     & $\mbox{D}^0$    & $q<0$ \\
$P_{5}$ &  $(1,0,0 )$      & $-\frac{1}{3}$   &  Saddle  & Saddle  & Saddle   & $\mbox{D}^-$ 	& $q=0$ \\
$L_{04}$&  $(s,-s,0)$      & $0$              &          & Sink    &          & $\mbox{D}^0$    & $q<0$ \\
\hline
\end{tabular}
\end{center}
\caption{Equilibrium points of the system \eqref{3De1}-\eqref{3De3} where ${\mu_{c}}=\sqrt{\frac{2}{3}}m$. The line of equilibria $L_{04}$ only exists when $\mu=\mu_{c}$ where $0<s<1$ and $P_{0}$ and $P_{4}$ are its endpoints. }\label{Table1}
\end{table}

\subsubsection{Stability of Equilibrium point $P_0$}

Evaluating the linearization matrix of the system \eqref{3De1}-\eqref{3De3} at $P_0$ gives us the following eigenvalues
\begin{eqnarray*}
&\lambda_{1}&=0,\\
&\lambda_{2,3}&= \pm \frac{\sqrt{6}}{2}\sqrt{\mu^2-{\mu_{c}}^2}.
\end{eqnarray*}
Note that, if $ \mu > {\mu_{c}}$ then $P_0$ is a saddle. But, if  $ \mu < {\mu_{c}}$ then  all the eigenvalues  have  zero real part which implies that the local qualitative behaviour at $P_0$ is not determined by its linearization.  However a perturbative solution near $P_0$  can be found, and fortunately an analysis of the first order solution is sufficient to determine the local stability of $P_0$ when $ \mu < {\mu_{c}}$.

We first introduce new scaled variables $(d,\phi,\psi)$ such that
\begin{equation}
D=\epsilon \left(d-\frac{\mu}{{\mu_{c}}}\phi\right),\quad \Phi=\epsilon \phi,\quad \Psi=\epsilon \psi,\label{RSC}
\end{equation}
where $\epsilon$ is assumed to be small, to determine a leading order approximation to the solution of the equations near $P_0$.  We note that the $\phi$ and $\psi$ variables that are employed in this subsection are not the original variables used to describe the scalar field and its derivative.  Using our new dependent variables \eqref{RSC}, and expanding \eqref{3De1}-\eqref{3De3} as a power series in $\epsilon$ we derive the following
\begin{eqnarray}
d^{\prime} &=&\frac{\epsilon}{3}\left(-d^2-2\psi^2+ \phi^2+2\frac{\mu}{{\mu_{c}}}d\phi-\frac{\mu^2}{{{\mu_{c}}}^2}\phi^2\right)+O(\epsilon^2), \nonumber\\
\phi^{\prime} &=& \frac{\sqrt{6}}{2}\mu_{c} \psi +O(\epsilon^2), \label{DS32}\\
\psi^{\prime} &=& \frac{\sqrt{6}}{2}\left(- \mu d -\mu_{c} \phi +\frac{\mu^2}{{\mu_{c}}}\phi\right)+ \epsilon\left(-d \psi+\frac{\mu}{{\mu_{c}}}\phi\psi \right)+O(\epsilon^2),\nonumber
\end{eqnarray}
where we kept only terms up to linear order in $\epsilon$.  To proceed with the construction of a perturbative solution, we employ the method of multiple scales \cite{hinch1991,kevorkian2013,nayfeh2000}.

In the method of multiple scales with two time scales, the original fast time $\tau$ and a second slow time $\eta=\epsilon\tau$, each dependent variable is expressed as
\begin{equation}
x \equiv    x(\tau, \eta)=    x_0 (\tau, \eta)+\epsilon    x_1 (\tau, \eta) + O(\epsilon^2) \label{CHinR}
\end{equation}
and using the chain rule, derivatives become expanded as
\begin{equation}
x^{\prime}=    x_{0\tau}+   \epsilon (x_{0\eta}      + x_{1\tau} )  + O(\epsilon^2).\label{var_exp}
\end{equation}
The $(\prime)$ indicates the ordinary derivative of the variable with respect to time $\tau$ while the subscripts $\tau$ and $\eta$ denote partial derivatives. A valid perturbative solution is obtained by ensuring that the solution remains bounded at all orders of $\epsilon$.

Using equation \eqref{CHinR} and \eqref{var_exp} for variables $(d,\phi,\psi)$ and substituting into  \eqref{DS32} and matching powers of $\epsilon$ yields the following system of partial differential equations for the zeroth order $[O(\epsilon^0)]$ terms
\begin{eqnarray}
d_{0\tau}    &=&  0, \nonumber\\
\phi_{0\tau} &=& \frac{\sqrt{6}}{2}\mu_{c}\psi_{0}, \\
\psi_{0\tau} &=&  \frac{\sqrt{6}}{2}\left(-\mu d_{0} - \mu_{c}\phi_{0}+\frac{\mu^2}{{\mu_{c}}}\phi_{0}\right),\nonumber
\label{L21}
\end{eqnarray}
and the following system of partial differential equations for the first order $[O(\epsilon^1)]$ terms
\begin{eqnarray}
d_{1\tau} &=&  \frac{1}{3}\left(-{d_{0}}^2-2{\psi_{0}}^2 + {\phi_{0}}^2 +2\frac{\mu}{{\mu_{c}}}d_{0}\phi_{0}-\frac{\mu^2}{{{\mu_{c}}}^2}{\phi_{0}}^2\right)-d_{0 \eta }, \nonumber\\
\phi_{1\tau}&=&  \frac{\sqrt{6}}{2}{\mu_{c}}\psi_{1}-\phi_{0 \eta},\label{L22}\\
\psi_{1\tau}&=&  \frac{\sqrt{6}}{2}\left(-\mu d_{1}-{\mu_{c}}\phi_{1} +\frac{\mu^2}{{\mu_{c}}}\phi_{1}\right)
+\left(-d_{0}\psi_{0}+\frac{\mu}{{\mu_{c}}}\phi_{0}\psi_{0}\right)-\psi_{0 \eta }. \nonumber
\end{eqnarray}
Solving the partial differential equations for the Zeroth order terms yields
\begin{eqnarray}
    d_{0}(\tau,\eta) &=& B(\eta), \nonumber\\
 \phi_{0}(\tau,\eta) &=& A(\eta)\cos(\lambda\tau-\Lambda(\eta))-\frac{\mu{\mu_{c}}}{{\mu_{c}}^2-\mu^2}B(\eta),\\
 \psi_{0}(\tau,\eta) &=& -\frac{\sqrt{6}\lambda}{3{\mu_{c}}}A(\eta)\sin(\lambda\tau-\Lambda(\eta)),\nonumber
\end{eqnarray}
where $\lambda=\frac{\sqrt{6}}{2}\sqrt{{\mu_{c}}^2-\mu^2}$ and $A(\eta)$, $B(\eta)$ and $\Lambda(\eta)$ are as yet undetermined functions of the slow time $\eta$.  Solving the partial differential equations for the first order terms, and restricting ourselves to only bounded solutions, determines a set of ordinary differential equations for the unknown functions $A(\eta)$, $B(\eta)$ and $\Lambda(\eta)$,
\begin{eqnarray}
       A_{\eta} &=&-\frac{1}{2}\frac{{\mu_{c}}^2}{{\mu_{c}}^2-\mu^2}AB, \nonumber\\
       B_{\eta} &=&-\frac{1}{3}\left(\frac{{\mu_{c}}^2}{{\mu_{c}}^2-\mu^2}B^2 + \frac{{\mu_{c}}^2-\mu^2}{2{\mu_{c}}^2}A^2\right),\label{Perturb_DE}\\
\Lambda_{\eta}  &=& 0\nonumber.
\end{eqnarray}

Therefore, in terms of the original variables the first term of the perturbative solution is
\begin{eqnarray*}
   D(\tau) &=& \epsilon\left(- \frac{\mu}{{\mu_{c}}}A(\eta)\cos(\lambda\tau-\Lambda(\eta))+\frac{{\mu_{c}}^2}{{\mu_{c}}^2-\mu^2}B(\eta) \right), \\
\Phi(\tau) &=& \epsilon\left(  A(\eta)\cos(\lambda\tau-\Lambda(\eta))-\frac{\mu{\mu_{c}}}{{\mu_{c}}^2-\mu^2}B(\eta)\right),\\
\Psi(\tau) &=& \epsilon\left(-\frac{\sqrt{{\mu_{c}}^2-\mu^2}}{{\mu_{c}}}A(\eta)\sin(\lambda\tau-\Lambda(\eta))\right),
\end{eqnarray*}
where the functions $A(\eta)$, $B(\eta)$ and $\Lambda(\eta)$ satisfy the differential equations \eqref{Perturb_DE}, and due to \eqref{Friedmann_Constraintfrw} are bounded by
\begin{equation}
B(\eta)^2\geq \frac{({\mu_{c}}^2-\mu^2)^2}{{\mu_{c}}^4}A(\eta)^2,
\end{equation}
where we note that if $B(\eta)\to 0$ then we also have $A(\eta)\to 0$.

We are interested in determining the asymptotic behaviour as $\tau \to \infty$.  We observe that the phase shift $\Lambda(\eta)$ is a constant and has no effect on the future dynamics.  The fast time $\tau$ essentially describes the oscillations of the scalar field, which to first order in $\epsilon$ has a period of $T=2\pi/\lambda$.  We note that the period of these oscillations $T \sim 1/\sqrt{{\mu_{c}}^2-\mu^2}$, gets longer as the strength of the coupling parameter $\mu$ is increased towards ${\mu_{c}}$.

We also observe that the amplitude of the oscillations $A(\eta)$, and the vertical shift $B(\eta)$ are functions of the slow time $\eta$ and consequently the amplitude and vertical shift drift slowly in comparison to the oscillatory changes. For initial values of $B(\eta)>0$ we see that both $A(\eta),B(\eta)\to 0$ as $\eta\to \infty$.  That is, the amplitude of the oscillations and the vertical shift both slowly decrease to zero, indicating that the point $P_0$ is stable when $\mu<{\mu_{c}}$.

\subsubsection{Stability of Equilibrium points in $\mbox{D}^-\cup\mbox{D}^0$}

Unfortunately, while we have an autonomous system of differential equations defined on a compact set, the system is not differentiable at any points in the invariant set $D^-\cup D^0$. We note that any equilibrium points in the invariant set $D^-\cup D^0$  represent asymptotic states in which $\theta \to \infty$.  In order to determine the local qualitative behaviour at these equilibrium points, we replace variable $D$ with
\begin{equation}
T=\frac{1}{\sqrt{1+\theta^2}}=\sqrt{1-D^2}.
\end{equation}
The evolution equations \eqref{DS1}-\eqref{DS3} become
\begin{eqnarray}
   T^{\prime} &=& -T\sqrt{1-T^2}\mathcal{X}, \\
\Psi^{\prime} &=& -\sqrt{1-T^2} \Psi- T \left( m \Phi+\sqrt{\frac{3}{2}}\mu \sqrt{1-T^2} \right) -\Psi \sqrt{1-T^2}\mathcal{X},\\
\Phi^{\prime} &=& m \Psi T-\sqrt{1-T^2} \Phi \mathcal{X},
\label{3Dmn012B}
\end{eqnarray}
with
\begin{equation}
\mathcal{X} = -\frac{1}{3}(1-T^2)-\frac{2}{3}\Psi^2+\frac{1}{3} \Phi^2 -\sqrt{\frac{3}{2}}\mu \Psi T.
\end{equation}
The value $D=1$ for the equilibrium points $P_{1,2,3,4,5}$ is simply replaced with $T=0$.  With this transformation we are able to locally determine the qualitative behaviour of each the equilibrium points in the invariant set $D^-\cup D^0$.

The eigenvalues of the points $P_1$ and $P_2$ are $1,\frac{4}{3}, 1$ which implies these points are generally sources.  The eigenvalues of the point $P_5$ is $\frac{1}{3},\frac{1}{3}, -\frac{2}{3}$ which implies this point is generally a saddle.  Further, the eigen-directions that span the $T=0$ invariant set, are associated with one positive and one negative eigenvalue.  Therefore this equilibrium point is a saddle in the $T=0$ set.

The eigenvalues of the point $P_3$ and $P_4$ are $0,-1, -\frac{2}{3}$ which implies that we cannot determine the general behaviour of this point without resorting to additional analysis.  However, the eigen-directions that span the $T=0$ invariant set, are associated with the two negative eigenvalues.  Therefore these equilibrium points are sinks in the $T=0$ set.

To complete the analysis of the qualitative behaviour near $P_3$ and $P_4$ we calculate the center manifold \cite{wiggins2003}.  In this case the center manifold is a one dimensional curve that must lie in the $\mbox{FRW}^0$ invariant set.  The center manifold for $P_3$ can be parameterized as
\begin{eqnarray}
T &=& T \\
\Phi &=& 1 - \left(\frac{1}{2} + \frac{3}{4}(\mu+{\mu_{c}})^2\right)T^2 + O(T^4)\\
\Psi &=& -\frac{\sqrt{6}}{2}(\mu+{\mu_{c}})T+\frac{3}{8}\sqrt{6}{\mu_{c}}(\mu+{\mu_{c}})^2T^3+O(T^4)
\end{eqnarray}
The leading order term of the dynamical system restricted to the center manifold reduces to
\begin{equation}
T'= \frac{3}{2}{\mu_{c}}(\mu+{\mu_{c}})T^3.
\end{equation}
Since $T'>0$ for $T>0$, $P_3$ is unstable along its center manifold. It is therefore a saddle in the full three dimensional phase space.

The center manifold for $P_4$ can be parameterized as
\begin{eqnarray}
T &=& T \\
\Phi &=& -1 + \left(\frac{1}{2} + \frac{3}{4}(\mu-{\mu_{c}})^2\right)T^2 + O(T^4)\\
\Psi &=& -\frac{\sqrt{6}}{2}(\mu-{\mu_{c}})T-\frac{3}{8}\sqrt{6}{\mu_{c}}(\mu-{\mu_{c}})^2T^3+O(T^4)
\end{eqnarray}
The leading order term of the dynamical system restricted to the center manifold reduces to
\begin{equation}
T'= -\frac{3}{2}{\mu_{c}}(\mu-{\mu_{c}})T^3.
\end{equation}
If $\mu<{\mu_{c}}$ then $T'>0$ for $T>0$ and $P_4$ is unstable along its center manifold.  However, if $\mu>{\mu_{c}}$ then $T'<0$ for $T>0$ and $P_4$ is stable along its center manifold. Therefore $P_4$ is a saddle when $\mu<{\mu_{c}}$ and a sink when $\mu>{\mu_{c}}$ in the full three dimensional phase space.

\subsubsection{The Bifurcation Value}

When $\mu={\mu_{c}}$, there is a line of equilibria given by $(D,\Phi,\Psi)=(s, -s,0)$ where $0<s<1$ having endpoints $P_0$ and $P_4$.  The eigenvalues of points on this line of equilibria are $0, -s, -\frac{2}{3}s$ which implies that points on this line are attractors when it exists. We observe that as the parameter $\mu$ increases towards its bifurcation value, ${\mu_{c}}$, the stability of the point $P_0$ is transferred to the point $P_4$ via this line of equilibria.

\subsubsection{Heteroclinic Sequences}

\begin{figure}
$\begin{array}{rl}
\setlength{\unitlength}{1cm}
\begin{picture}(8,8)
\thicklines
\put(0.9,1.8){$P_2$}
\put(1.5,1.9){\vector(1,0){2}}  
\put(1.55,2.0){\vector(1,1){1.9}}  
\put(1.5,2.1){\vector(1,2){1.9}}  
\put(0.9,6){$P_1$}
\put(1.5,6.1){\vector(1,0){2}}  
\put(1.55,6.0){\vector(1,-1){1.9}}  
\put(1.5,5.9){\vector(1,-2){1.9}}  

\put(3.6,6){$P_3$}
\put(4.2,6.1){\vector(1,-1){1.9}}  
\put(3.6,3.9){$P_5$}
\put(4.2,4.05){\vector(1,0){2}}  
\put(3.8,4.3){\vector(0,1){1.5}}  
\put(3.8,3.7){\vector(0,-1){1.5}}  
\put(3.6,1.8){$P_4$}
\put(4.2,2.0){\vector(1,1){1.9}}  
\put(6.3,3.9){$P_0$}
\put(1.1,1.6){\line(0,-1){0.5}}
\put(1.1,1.1){\line(1,0){5.4}}
\put(6.5,1.1){\vector(0,1){2.6}}
\put(1.1,6.6){\line(0,1){0.5}}
\put(1.1,7.1){\line(1,0){5.4}}
\put(6.5,7.1){\vector(0,-1){2.6}}
\put(3.4,0.2){$\mu<{\mu_{c}}$}
\end{picture}
&
\setlength{\unitlength}{1cm}
\begin{picture}(8,8)
\thicklines
\put(0.9,1.8){$P_2$}
\put(1.55,2.0){\vector(1,1){1.9}}  
\put(1.5,2.1){\vector(1,2){1.9}}  
\put(0.9,6){$P_1$}
\put(1.5,6.1){\vector(1,0){2}}  
\put(1.55,6.0){\vector(1,-1){1.9}}  
\put(1.5,5.9){\vector(1,-2){1.9}}  

\put(3.6,6){$P_3$}
\put(4.2,6.1){\vector(1,-1){1.9}}  
\put(3.6,3.9){$P_5$}
\put(4.2,4.05){\vector(1,0){2}}  
\put(3.8,4.3){\vector(0,1){1.5}}  
\put(3.8,3.7){\vector(0,-1){1.5}}  
\put(3.6,1.8){$P_0$}
\put(4.2,2.0){\vector(1,1){1.9}}  
\put(6.3,3.9){$P_4$}
\put(1.1,1.6){\line(0,-1){0.5}}
\put(1.1,1.1){\line(1,0){5.4}}
\put(6.5,1.1){\vector(0,1){2.6}}
\put(1.1,6.6){\line(0,1){0.5}}
\put(1.1,7.1){\line(1,0){5.4}}
\put(6.5,7.1){\vector(0,-1){2.6}}
\put(3.4,0.2){$\mu>{\mu_{c}}$}
\end{picture}
\end{array}$
\caption{The heteroclinic sequences indicating the past (sources are on the left-hand side), intermediate (middle) and late behaviour (sinks are on the right-hand side) for the negative curvature models.  The heteroclinic sequences for the flat models can be obtained by removing $P_5$ and any lines connected to it from the diagram above.}\label{fig1}
\end{figure}
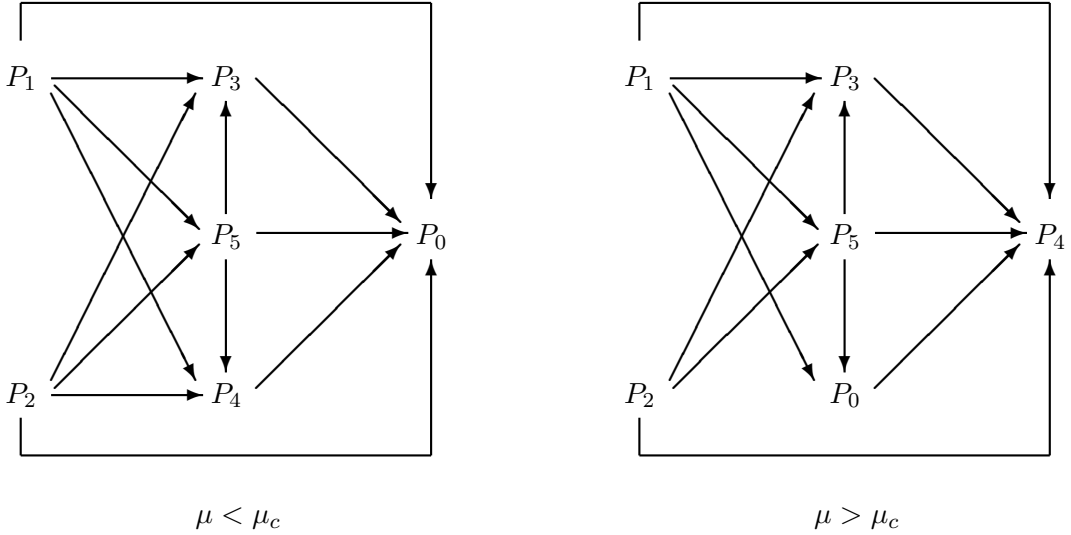

Very often one is not only interested in the past and future behaviour of a system of differential equations, but one is also interested in the intermediate behaviour of the system.  One technique to analyze the intermediate behaviour is to describe the heteroclinic sequences that are possible \cite{wainwright_ellis2005}.  We note that for every heteroclinic sequence there exists a set of orbits that are arbitrarily close to that sequence.  Figures \eqref{fig1} describes the possible heteroclinic sequences that are possible. Again we see how the intermediate and future behaviour changes as the parameter $\mu$ changes.

\subsection{Inflation and Accelerated Expansion}
As an indicator of the accelerated expansion we introduce the deceleration parameter
\begin{equation}
q \equiv -\frac{ a \ddot{a}}{\dot{a}^2}=- \left(3\frac{\dot{\theta}}{\theta^2}+1\right). \label{def_q}
\end{equation}
It follows that the deceleration parameter can also be expressed in terms of the normalized bounded variables as follows;
\begin{equation}
q= -\frac{1}{D^2} \left(-2 \Psi^2 +\Phi^2-3\sqrt{\frac{3}{2}} \mu \sqrt{1-D^2} \Psi \right).
\end{equation}
The sign of the deceleration parameter indicates the nature of the expansionary evolution. If $q > 0$, then the cosmological expansion is decelerating, while negative values of $q$ indicate an accelerating or inflationary dynamics. See Table \eqref{Table1} for a summary of the values of $q$ for each equilibrium point.

\subsection{Numerical Analysis}

\begin{figure}
$\begin{array}{rl}
\includegraphics[width=0.5\textwidth]{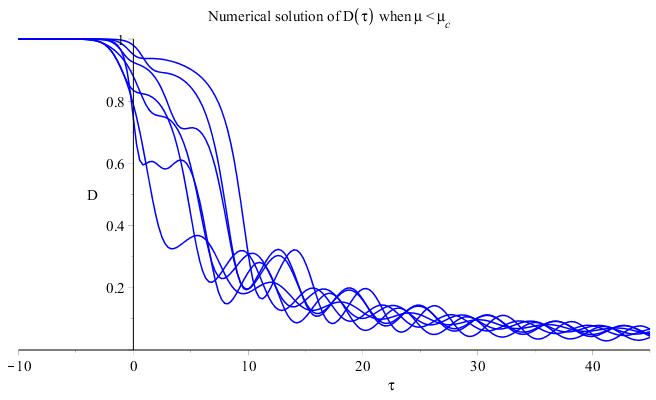} & \includegraphics[width=0.5\textwidth]{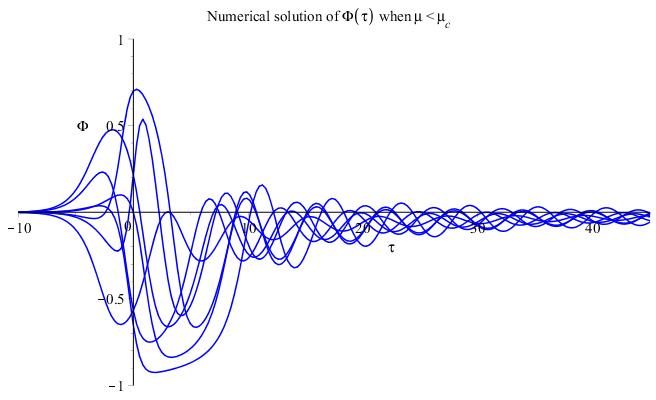} \\
\includegraphics[width=0.5\textwidth]{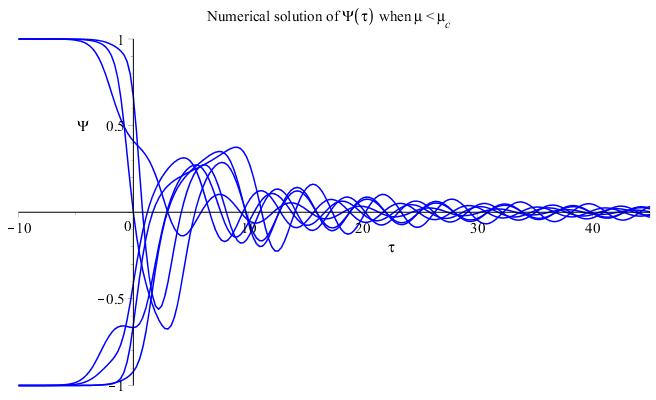}& \includegraphics[width=0.5\textwidth]{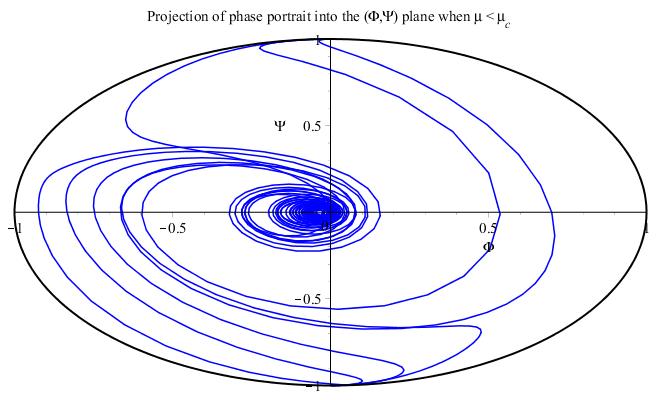}
\end{array}$
\caption{Numerical solutions of the system of differential equations \eqref{3De1}-\eqref{3De3} when $\mu<{\mu_{c}}$. Note how the amplitude of the oscillations in all the variables slowly decreases. Note the future asymptotic behaviour is  $(D,\Phi,\Psi) \to (0, 0, 0)$, $P_0$. There are two different possible past behaviours, one in which $(D,\Phi,\Psi)\to (1,0,1)$, $P_1$, and one in which $(D,\Phi,\Psi)\to (1,0,-1)$, $P_2$.}\label{FIG-FRW-MU-LESS}
\end{figure}

\begin{figure}
$\begin{array}{rl}
\includegraphics[width=0.5\textwidth]{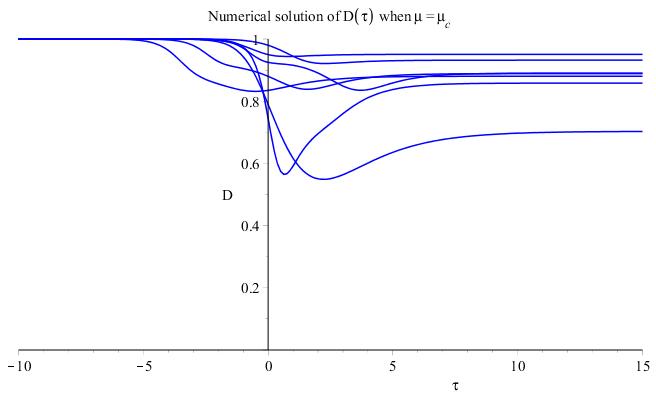} & \includegraphics[width=0.5\textwidth]{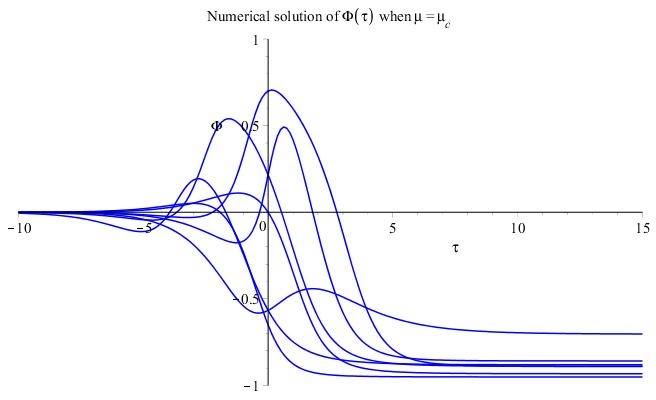} \\
\includegraphics[width=0.5\textwidth]{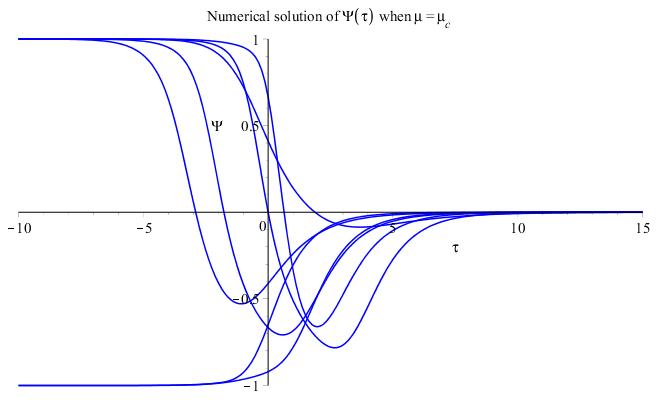}& \includegraphics[width=0.5\textwidth]{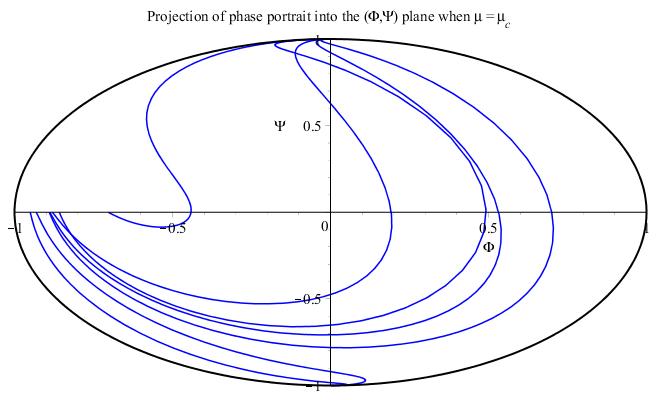}
\end{array}$
\caption{Numerical solutions of the system of differential equations \eqref{3De1}-\eqref{3De3} when $\mu={\mu_{c}}$, the bifurcation value. Note the future asymptotic behaviour $\Psi\to 0$ but both $D$ and $\Phi$ approach different future asymptotic states depending on their initial condition. The future asymptotic state is $L_{04}$.  There are two different possible past behaviours, one in which $(D,\Phi,\Psi)\to (1,0,1)$, $P_1$, and one in which $(D,\Phi,\Psi)\to (1,0,-1)$, $P_2$.}\label{FIG-FRW-MU-EQUAL}
\end{figure}

\begin{figure}
$\begin{array}{rl}
\includegraphics[width=0.5\textwidth]{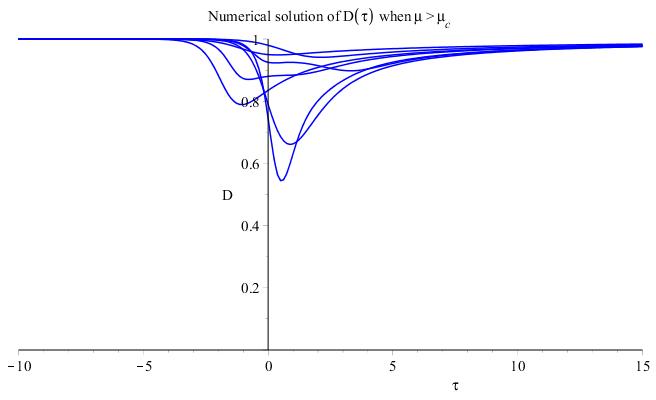} & \includegraphics[width=0.5\textwidth]{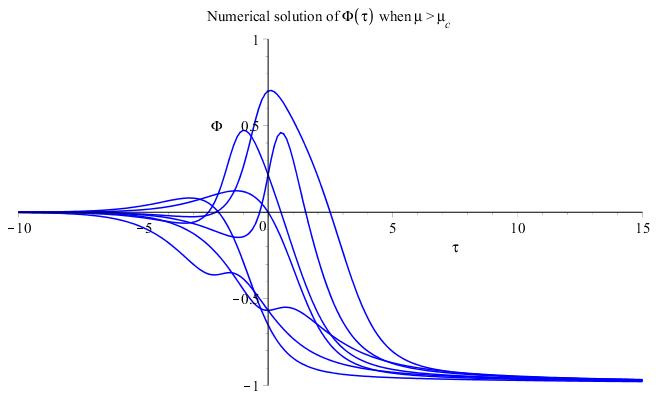} \\
\includegraphics[width=0.5\textwidth]{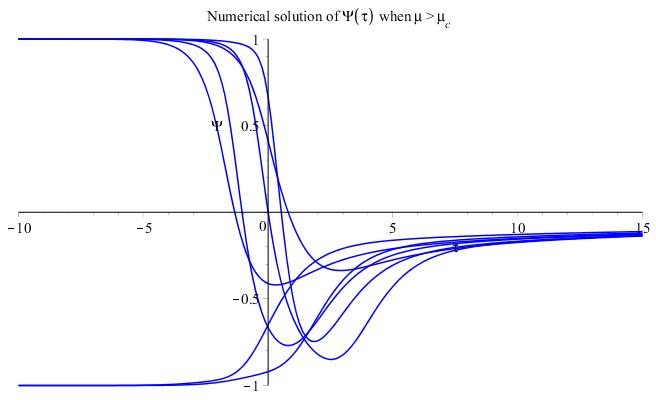}& \includegraphics[width=0.5\textwidth]{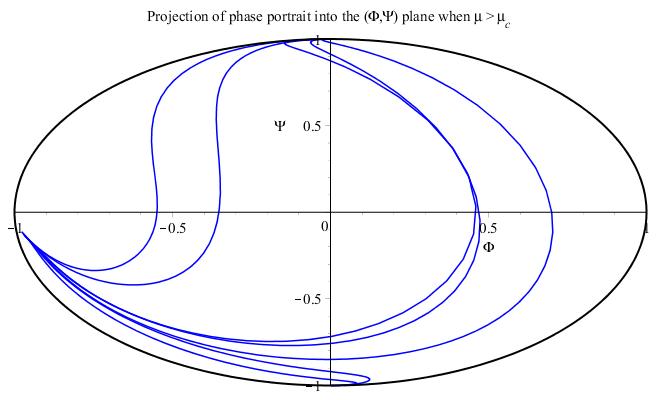}
\end{array}$
\caption{Numerical solutions of the system of differential equations \eqref{3De1}-\eqref{3De3} when $\mu>{\mu_{c}}$. Note the future asymptotic behaviour is  $(D,\Phi,\Psi) \to (1, 0, -1)$, $P_4$.  There are two different possible past behaviours, one in which $(D,\Phi,\Psi)\to (1,0,1)$, $P_1$, and one in which $(D,\Phi,\Psi)\to (1,0,-1)$, $P_2$.}\label{FIG-FRW-MU-GREAT}
\end{figure}

It is constructive to illustrate a few numerical solutions for the three different regimes of future asymptotic behaviour, $\mu<{\mu_{c}}$ [see Figure \eqref{FIG-FRW-MU-LESS}], $\mu={\mu_{c}}$ [see Figure \eqref{FIG-FRW-MU-EQUAL}], and $\mu>{\mu_{c}}$ [see Figure \eqref{FIG-FRW-MU-GREAT}]. In each case the integrations are done in the full 3-dimensional phase space. A few initial conditions are selected to show different past and future asymptotic behaviours.

\subsection{Observations}

In the spatially homogeneous and isotropic case analyzed here we observe that the past dynamics are independent of the strength of the coupling parameter $\mu$. We find in the zero curvature and the negative curvature models that there are two possible asymptotic behaviours to the past, one in which $(D,\Phi,\Psi)\to (1,0,1)$, $P_1$, and one in which $(D,\Phi,\Psi)\to (1,0,-1)$, $P_2$.  These past attractors represent a massless scalar field FRW model \cite{Coley:2003mj}.

We also observe that the future asymptotic state depends on the strength of the coupling parameter $\mu$.  For weak coupling of the Aether field to the scalar field, i.e, , $\mu<{\mu_{c}}=\sqrt{\frac{2}{3}}m$, the dynamics are similar to that found when there is no coupling of the Aether field to the scalar field, i.e., when $\mu=0$.  If $\mu<{\mu_{c}}$ then $P_0$ is the stable attractor: orbits oscillate and slowly decay in amplitude towards this final non-inflationary asymptotic state.

For strong coupling of the Aether field to the scalar field, $\mu>{\mu_{c}}$, the dynamics are very different.  When $\mu>{\mu_{c}}$ the scalar field does not come to rest at the minimum of the potential: the strength of the Aether interaction forces a different final equilibrium state.  If $\mu>{\mu_{c}}$ then in both the zero curvature and the negative curvature models we find that the stable attractor in these models changes from $P_0$ to $P_4$.  In this scenario, we find that the square of the scalar field and the square of the expansion scalar scale together as
\begin{equation}
\frac{m^2}{2}\phi^2 \sim \frac{1}{3}(1+\theta^2)
\end{equation}
and consequently grow without bound.  We also observe that all orbits (excepting for the exceptional orbits) experience some period of accelerating expansion as they evolve to their final asymptotic state, $P_4$.  In the next section we will add anisotropy to these models to determine if the addition of anisotropy changes these observations.


\section{Anisotropic Einstein-Aether Models coupled to a Scalar Field} \label{section3}

\subsection{A Class of Diagonal Bianchi Models}

In order to investigate Einstein-Aether cosmological models with a scalar field having a potential with interaction terms that contain both the expansion and shear of the aether vector requires a broader class of space-time geometries which include anisotropy.  For our purposes, the one parameter family of spatially homogeneous and anisotropic diagonal Bianchi type $VI_{h}$ space-times provides an arena to determine the effect of adding a shear interaction term into the scalar field potential in the Einstein-Aether models studied in the previous section.   The metric is assumed to have the form
\begin{equation}
ds^2=-dt^2+a(t)^2dx^2+b(t)^2e^{2(h-1)x}dy^2+c(t)^2e^{2x}dz^2.
\end{equation}
There are three special classes are worth mentioning: if the parameter $h=2$ then the metric is Bianchi type $V$ which has the negatively curved isotropic models (see previous section) as a subcase, if $h=1$ then it is Bianchi type $III$, and if $h=0$ then it is Bianchi type $VI_0$.

In a spatially homogeneous and isotropic cosmological model with comoving time, the aether vector field necessarily coincides with the rest frame defined by the Hubble expansion. Therefore in deviations from spatially homogeneity and isotropy, one could assume that the preferred frame for the aether approximately coincides with the cosmological rest frame defined by the Hubble expansion. However, if one relaxes only the isotropy assumption, then the aether vector would be tilted away from the hyper-surface normal of the spatially homogeneous hyper-surfaces which could allow for a richer set of dynamical behaviours. However, it has been argued in \cite{Alhulaimi:2013sha,Carruthers:2010ii,Kanno:2006ty} that during cosmological evolution, the two frames will come into alignment. Given these arguments, we will assume that the aether vector is aligned with the hyper-surface unit normal to the surfaces of homogeneity and is of the form $u^a=(1,0,0,0)$.

With the above assumptions on the metric and the aether vector, the vorticity and the acceleration of the aether vector are zero, and the covariant derivative
\begin{equation}
\nabla_b u_{a}=\sigma_{ab}+\frac{1}{3}\theta(g_{ab}+u_a u_b),
\end{equation}
is simply determined by the expansion scalar
\begin{equation}
\theta = \nabla_au^a = \frac{\dot a}{a}+\frac{\dot b}{b}+\frac{\dot c}{c},
\end{equation}
and the shear tensor
\begin{equation}
\sigma_{ab}=u_{(a;b)}-\frac{1}{3}\theta (g_{ab}+u_a u_b).
\end{equation}
The shear tensor has the form $\sigma^a_{\phantom{a}b}=\mbox{Diag}[0,\sigma_1,\sigma_2,-(\sigma_1+\sigma_2)]$ where
\begin{eqnarray*}
\sigma_1&=&\frac{1}{3}\left(2\frac{\dot a}{a}-\frac{\dot b}{b}-\frac{\dot c}{c}\right),\\
\sigma_2&=&\frac{1}{3}\left(-\frac{\dot a}{a}+2\frac{\dot b}{b}-\frac{\dot c}{c}\right).
\end{eqnarray*}
We note that the shear scalar is
\begin{equation}
\sigma^2=\frac{1}{2}\sigma_{ab}\sigma^{ab}=\sigma_1^2+\sigma_2^2+\sigma_1\sigma_2.
\end{equation}

With the definition of $T_{ab}^{\textsc{U}}$ in equation \eqref{T_ab_AE}, the effective energy density $\rho^{\textsc{U}}$, isotropic pressure, $p^{\textsc{U}}$, energy flux $q_{a}^{\phantom{a}\textsc{U}}$, and anisotropic stress $\pi_{\phantom{a}b}^{a\phantom{b}\textsc{U}}$   due to the aether field are
\begin{eqnarray}
\rho^{\textsc{U}} & = & -\frac{1}{3}c_\theta \theta^2 -2c_\sigma \sigma^2,\\
p^{\textsc{U}} & = & \frac{1}{3}c_\theta \theta^2 +\frac{2}{3}c_\theta \dot\theta-2c_\sigma \sigma^2,\\
q_{a}^{\phantom{a}\textsc{U}}&=&0,\\
\pi_{\phantom{a}b}^{a\phantom{b}\textsc{U}}& = &2c_\sigma (\dot\sigma^a_{\phantom{a}b}+\theta\sigma^a_{\phantom{a}b}),
\end{eqnarray}
Where new parameters $c_{\theta}= (c_1 +3c_2 +c_3)$ and $c_\sigma=c_1+c_3$, defined before in \cite{Coley:2015qqa,Latta:2016jix}, allow for some efficiencies in notation since the field equations are independent of any other linear combinations of the $c_i$.

The Einstein-aether field equations reduce to the following set of linearly independent equations
\begin{eqnarray}
0&=& -\frac{1}{3}(1+c_\theta)\theta^2+(1-2c_\sigma)\sigma^2+\frac{h^2-h+1}{a^2}+8\pi G\rho^{\textsc{M}}, \label{Friedmann-2}\\
0 &=& -(1+c_\theta)\dot\theta -\frac{1}{3}(1+c_\theta) \theta^2 -2(1-2c_\sigma)\sigma^2 -\frac{8\pi G}{2}(\rho^{\textsc{M}}+3p^{\textsc{M}} ),\label{Raychaudhuri-2}\\
0&=& -(h+1)\sigma_1+(h-2)\sigma_2-\frac{8\pi G}{a^2} q_{2}^{\phantom{a}\textsc{M}}\label{heat_flux}\\
0 &=& (1-2c_\sigma)\dot\sigma_1+(1-2c_\sigma)\theta\sigma_1 +\frac{(h-2)^2}{3a^2} -8\pi G\pi^\textsc{M}_1, \\
0 &=& (1-2c_\sigma)\dot\sigma_2+(1-2c_\sigma)\theta\sigma_2 +\frac{(h-2)(h+1)}{3a^2} -8\pi G\pi^\textsc{M}_2,
\end{eqnarray}
where there still exists the freedom to choose some appropriate units.  Without loss of generality, one can choose new units so that $\frac{8\pi G}{1+c_\theta}=1$, and a new parameter $C=\frac{1-2c_\sigma}{1+c_\theta}$, in which case the explicit dependence of the field equations on the aether parameter $c_\theta$ has again been eliminated.  Assuming that GR and Einstein Aether theory have equivalent PPN parameters and that there is stable positive energy modes but no vacuum \v{C}erenkov radiation imposes some constraints on the values of $c_\theta$ and $C$.  See Appendix \ref{Appendix} for details.

\subsection{The Scalar Field Potential}
We shall consider a class of quadratic scalar field potentials  of the form
\begin{equation}
V(\phi, \theta)=\frac{1}{2} m^2 \phi^2 + \mu \theta \phi+\nu\sigma\phi.\label{potential}
\end{equation}
In this case, from equation \eqref{scalar_T}, the effective energy density $\rho^{\textsc{M}}$, isotropic pressure $p^{\textsc{M}}$, energy flux $q_{a}^{\phantom{a}\textsc{M}}$, and anisotropic stress $\pi_{\phantom{a}b}^{a\phantom{b}\textsc{M}}$ due to the scalar field are
\begin{eqnarray}
\rho^{\textsc{M}} & = & \frac{1}{2}\dot\phi^2+ \frac{1}{2} m^2 \phi^2,\\
p^{\textsc{M}} & = & \frac{1}{2}\dot\phi^2- \frac{1}{2} m^2 \phi^2+\mu\dot\phi-\nu\phi\sigma,\\
q_{a}^{\phantom{a}\textsc{M}} & = & 0 \\
\pi_{\phantom{a}b}^{a\phantom{b}\textsc{M}}& = &
       \left(\frac{\nu\phi}{2\sigma}\right) \left(\dot\sigma^a_{\phantom{a}b}+\theta\sigma^a_{\phantom{a}b}\right)
 +\frac{d}{dt}{\left(\frac{\nu\phi}{2\sigma}\right)}  \sigma^a_{\phantom{a}b}.
\end{eqnarray}
where we define $\pi^\textsc{M}_i$ such that $\pi_{\phantom{a}i}^{i\phantom{b}\textsc{M}}=\mbox{Diag}[0,\pi^\textsc{M}_1,\pi^\textsc{M}_2,-(\pi^\textsc{M}_1+\pi^\textsc{M}_2)]$.
The final field equation comes from the Klein-Gordon equation which becomes
\begin{equation}
0=\ddot{\phi}+\theta\dot{\phi}+m^2\phi+\mu\theta+\nu\sigma.\label{KG-2}
\end{equation}

\subsection{The Dynamical System}

Given that there is no energy flux either from the existence of the aether field or from the non-minimal coupling of the aether to the scalar field potential, we are able to use equation \eqref{heat_flux} to rewrite all equations in terms of the anisotropy scalar $\sigma$:
\begin{equation}
\sigma_1 = \frac{h-2}{\sqrt{3}\sqrt{h^2-h+1}}\sigma, \qquad
\sigma_2 = \frac{h+1}{\sqrt{3}\sqrt{h^2-h+1}}\sigma.
\end{equation}
Hence, the final form of the Einstein-aether field equations and the Klein-Gordon equation yield the following dynamical system
\begin{eqnarray}
\dot\theta &=& -\frac{1}{3} \theta^2 -2C\sigma^2-\psi^2+\frac{m^2}{2}\phi^2-\frac{3}{2}\mu\psi+\frac{3}{2}\nu\phi\sigma,\label{DS2-1}\\
\dot\sigma&=&-\sigma\theta+\frac{\nu}{2C}\left(\psi+\theta\phi\right)\nonumber\\
&&\qquad+\frac{(h-2)}{\sqrt{3}\sqrt{h^2-h+1}}\frac{1}{C}
\left(\frac{1}{3}\theta^2-C\sigma^2-\frac{1}{2}\psi^2-\frac{m^2}{2}\phi^2\right),\\
\dot\phi&=&\psi\\
\dot\psi&=& -\theta\psi-m^2\phi-\mu\theta-\nu\sigma.\label{DS2-3}
\end{eqnarray}
with first integral
\begin{equation}
\frac{\theta^2}{3} = \frac{m^2}{2}\phi^2+\frac{1}{2}\psi^2+C\sigma^2+\frac{h^2-h+1}{a^2}, \label{Friedmann3}\\
\end{equation}

Equations \eqref{DS2-1}-\eqref{DS2-3}, therefore, yield a four dimensional dynamical system for the variables $(\theta,\sigma, \phi,\psi)$  depending on five parameters $(h, C, m, \mu, \nu)$ having a first integral given by equation \eqref{Friedmann3}. We restrict our analysis to the diagonal Bianchi type $V$ models $(h=2)$.  We note that even in this very simple anisotropic case when we compare the evolution equation for the shear with would would happen if the scalar field potential did not have an interaction term, we notice that the negatively curved FRW models are no longer an invariant subset of the Bianchi type $V$ system \eqref{DS2-1}-\eqref{DS2-3}.  Since the system of equations when $h=2$ is invariant under the transformations $(\mu,\sigma,\phi,\psi) \mapsto -(\mu,\sigma,\phi,\psi)$, and $(\nu,\sigma) \mapsto -(\nu,\sigma)$, we can without loss of generality, assume that both $\mu\geq 0$ and $\nu\geq0$.  Given that the phase space for the dynamical system defined in equations \eqref{DS2-1}-\eqref{DS2-3} with first integral \eqref{Friedmann3} is not  bounded, we employ dimensionless variables \cite{Coley:2003mj,wainwright_ellis2005} which will transform the system into an autonomous system of differential equations on a bounded phase space.

\subsection{Qualitative Analysis}

\subsubsection{Introducing Normalized Variables}

Introducing a time variable $\tau$
\begin{equation}
\frac{d \tau}{dt} \equiv \sqrt{1+\theta^2},
\end{equation}
and  normalized variables
\begin{eqnarray*}
D       &\equiv & \frac{\theta}{\sqrt{1+\theta^2}}\\
\Sigma  &\equiv & \frac{\sqrt{3} \sigma}{\sqrt{1+\theta^2}},\\
\Phi    &\equiv & \sqrt{\frac{3}{2}}\left(\frac{m \phi}{\sqrt{1+\theta^2}}\right),\\
\Psi    &\equiv & \sqrt{\frac{3}{2}}\left(\frac{\dot{\phi}}{\sqrt{1+\theta^2}}\right).
\end{eqnarray*}
then, the Bianchi type $V$ evolution equations in equations \eqref{DS2-1}-\eqref{DS2-3} when $h=2$  become
\begin{eqnarray}
        D^{\prime} &=& (1-D^2)\mathcal{X},\label{4De1}\\
     \Phi^{\prime} &=& m \sqrt{1-D^2}\Psi-D \Phi \mathcal{X}\label{4De2} \\
     \Psi^{\prime} &=& -D \Psi(1+\mathcal{X})- \sqrt{1-D^2} \Biggl[  m \Phi+\sqrt{\frac{3}{2}} \mu D+ \frac{\nu}{\sqrt{2}} \Sigma \Biggr],\label{4De3}\\
   \Sigma^{\prime} &=& - \Sigma D(1+  \mathcal{X}) + \frac{\nu}{ \sqrt{2}  C } \left( \Psi \sqrt{1-D^2} +\frac{D \Phi}{m}\right),\label{4De4}
\end{eqnarray}
where  the prime here indicates the differentiation with to respect to $\tau$.   $\mathcal{X}$  is given by the  expression
\begin{eqnarray}
    \mathcal{X} &=&  \frac{\dot{\theta}}{1+\theta^2}\nonumber\\
                &=&-\frac{1}{3}D^2-\frac{2}{3}\Psi^2+\frac{1}{3} \Phi^2 -\frac{2}{3} C \Sigma^2 -\sqrt{\frac{3}{2}} \mu  \sqrt{1-D^2} \Psi + \frac{\nu}{\sqrt{2} m}\Sigma\Phi.
    \label{X}
\end{eqnarray}
The Friedmann equation \eqref{Friedmann3} when $h=2$ becomes
\begin{equation}
  D^2-\Phi^2-    C \Sigma^2 -\Psi^2=\frac{9}{a^2 ( 1+ \theta^2)}. \label{FR32}
\end{equation}
where if $C\geq0$, then it follows that
\begin{equation}
0 \leq \Phi^2+\Psi^2 + C \Sigma^2  \leq D^2\leq 1.\label{Friedmann_Constraint}
\end{equation}
That is, $D, \Phi, \Psi, \Sigma $ are bounded if $C>0$ and the phase space is a bounded set.  Hence forward, we shall restrict our analysis to $C\geq0$ case only.

\subsubsection{Invariant Sets and Monotonic Functions}
The phase space can be subdivided into four disjoint invariant sets according to the curvature of the model and whether $D=1$ ($\theta\to\infty$) or not.  A superscript ``$-$'' indicates that points in this set represent negatively curved models, while a superscript ``$0$'' indicates a flat model. The invariant sets are
\begin{eqnarray*}
\mbox{A}^- &=& \{(D,\Phi,\Psi,\Sigma)|D<1,   \Phi^2+\Psi^2+C\Sigma^2  < D^2\},\\
\mbox{A}^0 &=& \{(D,\Phi,\Psi,\Sigma)|D<1,   \Phi^2+\Psi^2+C\Sigma^2  = D^2\},\\
\mbox{D}^-    &=& \{(D,\Phi,\Psi,\Sigma)|D=1,   \Phi^2+\Psi^2+C\Sigma^2  < 1\},\\
\mbox{D}^0    &=& \{(D,\Phi,\Psi,\Sigma)|D=1,   \Phi^2+\Psi^2+C\Sigma^2  = 1\}.
\end{eqnarray*}
The dimensions of which are 4, 3, 3, and 2, respectively.  The invariant set $\mbox{A}^-$ represents the Bianchi type $V$ models while the invariant set $\mbox{A}^0$ represents the Bianchi type $I$ models.  Again, the shear interaction term in the scalar field potential plays a significant role, since $\Sigma=0$ will be an invariant set only if $\nu=0$.

Similar to the isotropic case, if we define $\Lambda_3 = D^2- \Phi^2-\Psi^2-C\Sigma^2$ and $\Lambda_4=D^2-1$ then
\begin{eqnarray}
 \frac{\Lambda_3^{\prime}}{\Lambda_3} &=& -\frac{2}{3}  D  ( 3\mathcal{X}+1 ),\\
 \frac{\Lambda_4^{\prime}}{\Lambda_4} &=& 2D\mathcal{X}.
\end{eqnarray}
in which case the non-negative function $W=(\Lambda_3)^2(\Lambda_4)^2$ has the derivative
\begin{equation}
W^\prime = -\frac{4}{3}WD.
\end{equation}
Since $W> 0$ and $W^\prime<0$  in the set $\mbox{A}^-$ we can conclude that there are no periodic orbits in this 4-dimensional invariant set. This also implies that there are no equilibrium points in the set $\mbox{A}^-$, and any equilibrium points of the autonomous system of differential equations \eqref{4De1}-\eqref{4De4} will lie in the lower dimensional invariant sets $\mbox{A}^0$, $\mbox{D}^-$  or $\mbox{D}^0$.

\subsubsection{Equilibrium Points}
The equilibrium points for  the system \eqref{4De1}-\eqref{4De4} with the value of $\mathcal{X}$  and their stability are summarized in Table \eqref{Table2}. Comparing the equilibrium points in Table \eqref{Table2} with those found in Table \eqref{Table1} we see that the points $P_0$ and $P_5$ represent the same equilibrium states in both tables. The points $P_1$ and $P_2$ represents the same equilibrium state in both tables, however, in Table \eqref{Table2} they are actually two special (isotropic) points on a non-isolated circle of generally non-isotropic equilibria given by $C^*$.  The points $P_3$ and $P_4$ in Table \eqref{Table2} reduce to $P_3$ and $P_4$ in Table \eqref{Table1} when $\nu=0$.  Further, the dynamical behaviour and stability is analogous to the stability of $P_3$ and $P_4$ in Table \eqref{Table1}. Similar to the isotropic case, there is a line of equilibria $L_{04}$ when $\mu=\mu_{c\nu}$ that connects $P_{0}$ and $P_4$.  We note that the local stability of the equilibrium points depends on the bifurcation value ${\mu_{c\nu}}^2={\mu_c}^2 +\frac{\nu^2}{3C}$. Recall ${\mu_{c}}=\sqrt{\frac{2}{3}}m$ is the bifurcation value found in the isotropic case studied earlier, and so since $C$ and $\nu$ are both positive, the bifurcation value for the anisotropic case, is always a bit larger than in the isotropic case. Further, by choosing smaller and smaller values of $C$, one can increase the value of bifurcation value $\mu_{c\nu}$.

\begin{table}
\begin{center}
\begin{tabular}{|c|c|c|l|l|l|c|c|}
\hline
Pt      & $(D,\Phi,\Psi,\Sigma)$  & $\mathcal{X}$   &\multicolumn{3}{|c|}{Stability}                          & Invariant      & $q$  \\ \cline{4-6}
 	
 	    &                        &                  & $\mu < \mu_{c\nu}$ & $\mu=\mu_{c\nu}$    &$\mu> \mu_{c\nu}$   & Set         &      \\
\hline
$P_{0}$ &  $(0,0, 0,0)$          & $0$              &	 Sink          & Sink             & Saddle            & $\mbox{A}^0$      & DNE \\
$P_{1}$ &  $(1, 0,1,0)$          & $-1$             &  Source	       & Source           & Source            & $\mbox{D}^0$      & $q>0$ \\
$P_{2}$ &  $(1,0,-1,0)$          & $-1$             &  Source	       & Source           & Source            & $\mbox{D}^0$      & $q>0$ \\
$P_{3}$ &  $\left(1,\frac{{\mu_{c}}}{\mu_{c\nu}}, 0,\frac{\nu}{\sqrt{3}C} \frac{1}{\mu_{c\nu}}\right)$
                                & $0$              &  Saddle	       & Saddle           & Saddle            & $\mbox{D}^0$      & $q<0$ \\
$P_{4}$ &  $\left(1,-\frac{{\mu_{c}}}{\mu_{c\nu}},0,-\frac{\nu}{\sqrt{3}C} \frac{1}{\mu_{c\nu}}\right)$
                                & $0$              &  Saddle	       & Sink             & Sink              & $\mbox{D}^0$      & $q<0$ \\
$P_{5}$ &  $(1,0,0,0 )$         & $-\frac{1}{3}$   &  Saddle          & Saddle            & Saddle            & $\mbox{D}^-$ 	  & $q=0$ \\
$C^{*}$ &   $(1,0,\sin(u),\frac{1}{\sqrt{C}}\cos(u))$
                                & $-1$             &  Source          &  Source           &  Source           & $\mbox{D}^0$      & $q<0$ \\
$L_{04}$&  $(s,-\frac{{\mu_{c}}}{\mu_{c\nu}}s,0,-\frac{\nu}{\sqrt{3}C}\frac{1}{\mu_{c\nu}}s)$
                                & $-1$             &                  & Sink              &                    &$\mbox{A}^0$   & $q<0$\\
                                \hline
\end{tabular}
\end{center}
\caption{Equilibrium points of the system \eqref{4De1}-\eqref{4De4} where ${\mu_{c\nu}}^2={\mu_c}^2 +\frac{\nu^2}{3C}$. For the circle of non-isolated equilibria $C^{*}$, $u\in [-\pi,\pi)$ where we note that $P_{1}$ and $P_{2}$ are actually the isotropic equilibrium points on this circle. The line of equilibria $L_{04}$ only exists when $\mu=\mu_{c\nu}$ where $0< s<1 $ and $P_{0}$ and $P_4$ are as its endpoints.}\label{Table2}
\end{table}

\subsubsection{Stability of Equilibrium Point $P_{0}$}

Evaluating the linearization matrix of the system \eqref{4De1}-\eqref{4De4} at $P_0$ gives us the following eigenvalues
\begin{eqnarray}
&\lambda_{1,2}&=0,\\
&\lambda_{3,4}&= \pm \frac{\sqrt{6}}{2}\sqrt{\mu^2-{\mu_{c \nu}} ^2}.
\end{eqnarray}
Note that, if $ \mu >\mu_{c \nu} $ then $P_0$ is a saddle. But, if  $ \mu < \mu_{c \nu} $  then  all the eigenvalues  have  zero real part which implies that the local qualitative behaviour at $P_0$ is not determined by its linearization.  However a perturbative solution near $P_0$  can be found, and fortunately an analysis of the first order solution is sufficient to determine the local stability of $P_0$ when $ \mu  < \mu_{c \nu} $ .

We first introduce new scaled variables $(d,\phi,\psi,\sigma)$ such that
\begin{equation}
D=\epsilon \left(d-\frac{\mu}{\mu_c}\phi\right),\quad \Phi=\epsilon \phi,\quad \Psi=\epsilon \psi, \quad \Sigma= \epsilon \left(\sigma +\frac{\nu}{\sqrt{3} C {\mu_{c}}} \phi\right)\label{RSC2}
\end{equation}
where $\epsilon$ is assumed to be small, to determine a leading order approximation to the solution of the equations near $P_0$.  We note that the $\phi,\psi$ and $\sigma$ variables that are employed in this subsection are not the original variables used to describe the scalar field and its derivative.  Using our new dependent variables \eqref{RSC2}, and expanding \eqref{4De1}-\eqref{4De4} as a power series in $\epsilon$ we derive the following
\begin{eqnarray}
d^{\prime} &=&\frac{\epsilon}{3}\left(-d^2-2\psi^2+ \phi^2+2\frac{\mu}{{\mu_{c}}}d\phi-\frac{\mu^2}{{{\mu_{c}}}^2}\phi^2 +\frac{ \nu^2 }{ 3 C {{\mu_{c}}}^2} \phi^2 -2 C \sigma^2 -\frac{\sqrt{3} \nu }{ 3 {\mu_{c}}} \phi \sigma \right)+O(\epsilon^2), \nonumber\\
\phi^{\prime} &=& \frac{\sqrt{6}}{2}{\mu_{c}} \psi +O(\epsilon^2), \label{DS322}\\
\psi^{\prime} &=& \frac{\sqrt{6}}{2}\left(- \mu d -{\mu_{c}} \phi +\frac{\mu^2}{{\mu_{c}}}\phi - \frac{\nu}{ \sqrt{3}} \sigma -\frac{ \nu^2}{ 3 C {\mu_{c}}} \phi \right)+ \epsilon\left(-d \psi+\frac{\mu}{{\mu_{c}}}\phi\psi \right)+O(\epsilon^2),\nonumber \\
\sigma^{\prime} & =&  \epsilon \left( -\sigma d +\frac{ \mu}{{\mu_{c}}} \sigma \phi\right).\nonumber
\end{eqnarray}
where we kept only terms up to linear order in $\epsilon$.  To proceed with the construction of a perturbative solution, we employ the method of multiple scales \cite{hinch1991,kevorkian2013,nayfeh2000}.

Using equation \eqref{CHinR} and \eqref{var_exp} for variables $(d,\phi,\psi,\sigma)$ and substituting into  \eqref{DS322} and matching powers of $\epsilon$ yields the following system of partial differential equations for the zeroth order $[O(\epsilon^0)]$ terms
\begin{eqnarray}
d_{0\tau}    &=&  0, \nonumber\\
\phi_{0\tau} &=& \frac{\sqrt{6}}{2}{\mu_{c}}\psi_{0}, \label{L212} \\
\psi_{0\tau} &=&  \frac{\sqrt{6}}{2}\left(- \mu d_{0} -{\mu_{c}} \phi_{0} +\frac{\mu^2}{{\mu_{c}}}\phi_{0} - \frac{\nu}{ \sqrt{3}} \sigma_{0} -\frac{ \nu^2}{ 3 C {\mu_{c}}} \phi_{0} \right),\nonumber \\
\sigma_{0\tau}    &=&  0 \nonumber
\end{eqnarray}
and the following system of partial differential equations for the first order $[O(\epsilon^1)]$ terms
\begin{eqnarray}
d_{1\tau} &=&  \frac{1}{3}\left(-{d_{0}}^2-2{\psi_{0}}^2+ {\phi_{0}}^2+2\frac{\mu}{{\mu_{c}}}d_{0} \phi_{0}-\frac{\mu^2}{{{\mu_{c}}}^2}{\phi_{0}}^2 +\frac{ \nu^2 }{ 3 C {{\mu_{c}}}^2} {\phi_{0}}^2 -2 C {\sigma_{0}}^2 -\frac{\sqrt{3} \nu }{ 3 {\mu_{c}}} \phi_{0} \sigma_{0} \right)-d_{0 \eta }, \nonumber\\
\phi_{1\tau}&=&  \frac{\sqrt{6}}{2}{\mu_{c}}\psi_{1}-\phi_{0 \eta},\label{L222}\\
\psi_{1\tau}&=&  \frac{\sqrt{6}}{2}\left(- \mu d_{1} -{\mu_{c}} \phi_{1} +\frac{\mu^2}{{\mu_{c}}}\phi_{1} - \frac{\nu}{ \sqrt{3}} \sigma_{1} -\frac{ \nu^2}{ 3 C {\mu_{c}}} \phi_{1} \right)
+\left(-d_{0}\psi_{0}+\frac{\mu}{{\mu_{c}}}\phi_{0}\psi_{0}\right)-\psi_{0 \eta }، \nonumber
\\
\sigma_{1\tau} &=& \left(-d_{0} \sigma_{0} +\frac{\mu}{{\mu_{c}}} \sigma_{0} \phi_{0}\right) -\sigma_{0 \eta }.\nonumber
\end{eqnarray}
Solving the partial differential equations for the Zeroth order terms yields

\begin{eqnarray}
    d_{0}(\tau,\eta) &=& B(\eta), \nonumber\\
 \phi_{0}(\tau,\eta) &=& A(\eta)\cos(\lambda\tau-\Lambda(\eta))-\frac{{\mu_{c}}}{({\mu_{c \nu }}^2 -\mu^2)} \left(\mu B(\eta)+\frac{\nu}{\sqrt{3}} S(\eta)\right),\\
 \psi_{0}(\tau,\eta) &=& -\frac{\sqrt{6}\lambda}{3{\mu_{c}}}A(\eta)\sin(\lambda\tau-\Lambda(\eta)),\nonumber \\
 \sigma_{0}(\tau,\eta) & =& S(\eta).\nonumber
\end{eqnarray}
where $\lambda=\frac{\sqrt{6}}{2}\sqrt{{\mu_{c \nu }}^2-\mu^2}$ and $A(\eta)$, $B(\eta), S(\eta)$ and $\Lambda(\eta)$ are as yet undetermined functions of the slow time $\eta$.  Solving the partial differential equations for the first order terms, and restricting ourselves to only bounded solutions, determines a set of ordinary differential equations for the unknown functions $A(\eta)$, $B(\eta), S(\eta)$ and $\Lambda(\eta)$.  If we replace $B(\eta)$ with the linear combination $${\bar B}(\eta)={\mu_{c\nu}}^2B(\eta)+\mu\frac{\nu}{\sqrt{3}}S(\eta)$$ then the resulting set of differential equations become
\begin{eqnarray}
A_{\eta}              &=&-\frac{1}{2} \frac{1}{{\mu_{c \nu }}^2-\mu^2} A {\bar B}, \nonumber\\
{\bar B}_{\eta}       &=&-\frac{1}{3} \left(\frac{1}{{\mu_{c\nu}}^2-\mu^2}{\bar B}^2 + \frac{{\mu_{c\nu}}^2({\mu_{c\nu}}^2-\mu^2)}{2{{\mu_{c}}}^2}A^2 + 2{\mu_{c}} C S^2 \right),\nonumber\\
S_{\eta}              &=&-            \frac{1}{{\mu_{c \nu }}^2-\mu^2} S {\bar B}, \nonumber\\
\Lambda_{\eta}        &=& 0. \label{Perturb_DE2}
\end{eqnarray}

Therefore, in terms of the original variables the first term of the perturbative solution is
\begin{eqnarray*}
   D(\tau) &=& \epsilon\left[ B(\eta) - \frac{\mu}{{\mu_{c}}}A(\eta)\cos(\lambda\tau-\Lambda(\eta))
   +\frac{\mu}{({\mu_{c\nu}}^2-\mu^2)} \left(\mu B(\eta)+\frac{\nu}{\sqrt{3}} S(\eta)\right) \right], \\
\Phi(\tau) &=& \epsilon\left[ A(\eta)\cos(\lambda\tau-\Lambda(\eta))
    -\frac{{\mu_{c}}}{({\mu_{c\nu}}^2-\mu^2)} \left(\mu B(\eta)+\frac{\nu}{\sqrt{3}} S(\eta)\right) \right],\\
\Psi(\tau) &=& \epsilon\left[-\frac{\sqrt{{\mu_{c \nu }}^2-\mu^2}}{{\mu_{c}}}A(\eta)\sin(\lambda\tau-\Lambda(\eta))\right],\\
\Sigma( \tau) &=& \epsilon \left[ S(\eta)+\frac{\nu}{ \sqrt{3} C {\mu_{c}}}   A(\eta)\cos(\lambda\tau-\Lambda(\eta))
    -\frac{\nu}{\sqrt{3}C({\mu_{c\nu}}^2-\mu^2)} \left(\mu B(\eta)+\frac{\nu}{\sqrt{3}} S(\eta)\right)\right],
\end{eqnarray*}
where the functions $A(\eta)$, $B(\eta)$, $S(\eta)$, and $\Lambda(\eta)$ satisfy the differential equations \eqref{Perturb_DE2}, and due to \eqref{Friedmann_Constraint} are bounded by
\begin{equation}
{\bar B}(\eta)^2 \geq \left(\frac{{\mu_{c\nu}}^2-\mu^2}{{{\mu_{c}}}^2}\right)  A(\eta)^2  + \frac{{{\mu_{c}}}^2}{{\mu_{c\nu}}^2}CS(\eta)^2,
\end{equation}
where we note that if ${\bar B}(\eta)\to 0$ then we also have $A(\eta)\to 0$ and $S(\eta)\to 0$ which then also implies that $B(\eta)\to 0$.

We are interested in determining the asymptotic behaviour as $\tau \to \infty$.  We observe that the phase shift $\Lambda(\eta)$ is a constant and has no effect on the future dynamics.  The fast time $\tau$ essentially describes the oscillations of the scalar field, which to first order in $\epsilon$ has a period of $T=2\pi/\lambda$.  We note that the period of these oscillations $T \sim 1/\sqrt{{\mu_{c \nu }}^2-\mu^2}$, gets longer as the strength of the coupling parameter $\mu$ is increased.

We also observe that the amplitude of the oscillations $A(\eta)$, and the vertical shift $B(\eta)$ and the shear term $S(\eta)$ are functions of the slow time $\eta$ and consequently the amplitude, vertical shift, and shear term drift slowly in comparison to the oscillatory changes. For initial values of ${\overline B}(\eta)>0$ we see that $A(\eta),B(\eta),S(\eta) \to 0$ as $\eta\to \infty$.  That is, the amplitude of the oscillations, the vertical shift, and the shear term all slowly decrease to zero, indicating that the point $P_0$ is stable when $\mu< {\mu_{c \nu }}.$

\subsubsection{Stability of Equilibrium Points in $\mbox{D}^-\cup\mbox{D}^0$}
Unfortunately, while  we have an autonomous system of differential equations defined on a compact set, the system is not differentiable  at any points in the invariant set $\mbox{D}^-\cup\mbox{D}^0$. In order to determine the local behaviour at these  points, we replace variable $ D $ with
\begin{equation}
 T=\frac{1}{ \sqrt{1+ \theta^2}}= \sqrt{1-D^2}.
\end{equation}
The evolution  equations \eqref{4De1}-\eqref{4De4} become
\begin{eqnarray}
T^{\prime}    &=& -T \sqrt{1-T^2} \mathcal{X},\quad \\
\Phi^{\prime} &=& m T\Psi-\sqrt{1-T^2} \Phi \mathcal{X} , \quad \\
\Psi^{\prime} &=& -\sqrt{1-T^2} \Psi(1+\mathcal{X})- T \left(  m \Phi+\sqrt{\frac{3}{2}} \mu \sqrt{1-T^2}+ \frac{\nu}{\sqrt{2}} \Sigma \right), \\
\Sigma^{\prime}&=& - \Sigma \sqrt{1-T^2}(1+  \mathcal{X}) + \frac{\nu}{ \sqrt{2}  C} \left( \Psi T +\frac{\sqrt{1-T^2} \Phi}{m}\right),
\end{eqnarray}
with
\begin{equation}
    \mathcal{X}=-\frac{1}{3}(1-T^2)-\frac{2}{3}\Psi^2+\frac{1}{3} \Phi^2 -\frac{2}{3}  C \Sigma^2 -\sqrt{\frac{3}{2}} \mu  T\Psi +\sqrt{\frac{3}{2}} \frac{\nu}{m}\Sigma\Phi.
    \label{X3}
\end{equation}
The value $D=1$ for the equilibrium points $P_{1}$ $P_{2}$, $P_{3}$, $P_{4}$, $P_{5}$ and $C^*$ is simply replaced with $T=0$. With this transformation we are able to locally determine the qualitative behaviour of each the equilibrium points.

The eigenvalues of the linearization at the equilibrium points $P_1$, $P_2$, and $C^*$ are $ 0, 1,1, \frac{4}{3}$ which implies these points are unstable and are sources. The zero eigenvalue indicates the non-isolated nature of the circle of equilibria $C^*$.

The eigenvalues of the linearization at the equilibrium point $P_{5}$  is $ \frac{1}{3}, -\frac{2}{3}, \frac{1}{3}, -\frac{2}{3}$ which implies that  $P_{5}$ is  generally saddle (unstable). Further, the eigen-directions that span the $T=0$  invariant set, are associated with one positive and two negative eigenvalues. Therefore this equilibrium point is a saddle within the  $T=0$ invariant set.

The eigenvalues of the linearization at the equilibrium points $ P_{3}$ and $P_{4}$ are $0, -1,-1, -\frac{2}{3}$ which implies that we cannot  determine the general behaviour of these points without resorting to additional analysis. However, the eigen-directions that span the $T=0$ invariant set, are associated with the three negative eigenvalues. Therefore, these equilibrium points are sinks in the $T=0$ set. One method to complete the determination of the general behaviour near $P_3$ and $P_4$ is to calculate the center manifold \cite{wiggins2003}.

The center manifold for $P_3$ can be parameterized as
\begin{eqnarray}
T &=& T \\
\Phi &=& \frac{{\mu_{c}}}{\mu_{c\nu}}-\frac{\mu}{\mu_{c\nu}}\left(1+\frac{3}{4}(\mu+\mu_{c\nu})^2\right)T^2+O(T^4),
\\
\Psi &=&  -\frac{\sqrt{6}}{2}(\mu+\mu_{c\nu})T+\frac{\sqrt{6}}{2}\mu_{c\nu}\left(\frac{1}{2}+\frac{3}{4}(\mu+\mu_{c\nu})^2\right)T^3+O(T^4),
\\
\Sigma & = & \frac{\nu}{\sqrt{3}C}\frac{1}{{\mu_{c\nu}}}-\frac{\nu}{\sqrt{3}C}\frac{1}{{\mu_{c\nu}}}\left(1+\frac{3}{4}(\mu+{\mu_{c\nu}})^2\right)T^2+ O(T^4)
\end{eqnarray}
The leading order term of the dynamical system restricted to the center manifold reduces to
\begin{equation}
T'= \frac{3}{2}{\mu_{c\nu}}(\mu+{\mu_{c\nu}}) T^3.
\end{equation}
Since $T'>0$ for $T>0$, $P_3$ is unstable along its center manifold. Therefore $P_3$ is a saddle within the full four dimensional phase space.

The center manifold for $P_4$ can be parameterized as
\begin{eqnarray}
T &=& T \\
\Phi &=& -\frac{{\mu_{c}}}{{\mu_{c\nu}}}+\frac{\mu}{{\mu_{c\nu}}}\left(1+\frac{3}{4}(\mu-{\mu_{c\nu}})^2\right)T^2+O(T^4),
\\
\Psi &=&  -\frac{\sqrt{6}}{2}(\mu-{\mu_{c\nu}})T-\frac{\sqrt{6}}{2}{\mu_{c\nu}}\left(\frac{1}{2}+\frac{3}{4}(\mu-{\mu_{c\nu}})^2\right)T^3+O(T^4),
\\
\Sigma & = & -\frac{\nu}{\sqrt{3}C}\frac{1}{{\mu_{c\nu}}}+\frac{\nu}{\sqrt{3}C}\frac{1}{{\mu_{c\nu}}}\left(1+\frac{3}{4}(\mu-{\mu_{c\nu}})^2\right)T^2+ O(T^4)
\end{eqnarray}
The leading order term of the dynamical system restricted to the center manifold reduces to
\begin{equation}
T'= -\frac{3}{2}{\mu_{c\nu}}(\mu-{\mu_{c\nu}}) T^3.
\end{equation}
If $ \mu > {\mu_{c \nu}}$ then $T'<0$ and $P_4$ is stable along its center manifold. It is therefore a sink when $ \mu > {\mu_{c \nu}}$, in the full four dimensional phase space and a saddle otherwise.

\subsubsection{The Bifurcation Value}

If $\mu={\mu_{c\nu}}$ then there is a line of equilibria given by $(D,\Phi,\Psi,\Sigma)=(s,-\frac{{\mu_{c}}}{{\mu_{c\nu}}}s,0,-\frac{\nu}{\sqrt{3}C}\frac{1}{{\mu_{c\nu}}}s)$ where $0<s<1$ having endpoints $P_0$ and $P_4$.  The eigenvalues of points on this line of equilibria are $0, -s,-s, -\frac{2}{3}s$ which implies that points on this line are attractors when it exists. We observe that as the parameter $\mu$ increases towards its bifurcation value, ${\mu_{c\nu}}$, the stability of the point $P_0$ is transferred to the point $P_4$ via this line of equilibria.

\subsubsection{Heteroclinic Sequences}

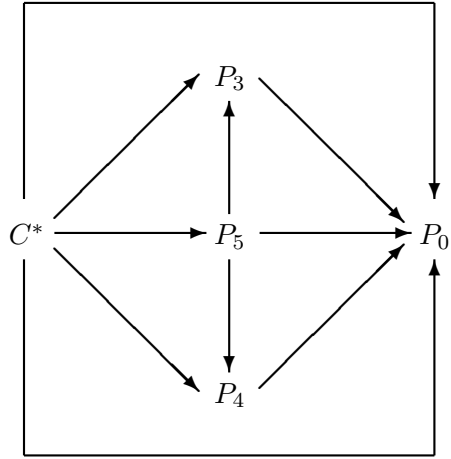
\begin{figure}
\setlength{\unitlength}{1cm}
\begin{center}
\begin{picture}(8,8)
\thicklines
\put(0.9,3.9){$C^*$}
\put(1.5,4.25){\vector(1,1){1.9}}  
\put(1.5,4.05){\vector(1,0){2}}  
\put(1.5,3.85){\vector(1,-1){1.9}}  
\put(3.6,6){$P_3$}
\put(4.2,6.1){\vector(1,-1){1.9}}  
\put(3.6,3.9){$P_5$}
\put(4.2,4.05){\vector(1,0){2}}  
\put(3.8,4.3){\vector(0,1){1.5}}  
\put(3.8,3.7){\vector(0,-1){1.5}}  
\put(3.6,1.8){$P_4$}
\put(4.2,2.0){\vector(1,1){1.9}}  
\put(6.3,3.9){$P_0$}
\put(1.1,3.7){\line(0,-1){2.6}}
\put(1.1,1.1){\line(1,0){5.4}}
\put(6.5,1.1){\vector(0,1){2.6}}
\put(1.1,4.5){\line(0,1){2.6}}
\put(1.1,7.1){\line(1,0){5.4}}
\put(6.5,7.1){\vector(0,-1){2.6}}
\end{picture}
\end{center}
\caption{The heteroclinic sequences indicating the past (sources are on the left-hand side), intermediate (middle) and late behaviour (sinks are on the right-hand side) when $\mu<{\mu_{c\nu}}$ for the anisotropic Bianchi type $V$ models.  The heteroclinic sequences for the flat Bianchi type$I$ models can be obtained by removing $P_5$ and any lines connected to it from the diagram above.  The heteroclinic sequences for the anisotropic Bianchi type $V$ models when $\mu>{\mu_{c\nu}}$ can be obtained by simply swapping $P_0$ and $P_4$.}\label{figB1}
\end{figure}

Very often one is not only interested in the past and future behaviour of a system of differential equations, but one is also interested in the intermediate behaviour of the system.  One technique to analyze the intermediate behaviour is to describe the heteroclinic sequences that are possible \cite{wainwright_ellis2005}.  We note that for every heteroclinic sequence there exists a set of orbits that are arbitrarily close to that sequence.  Figure \eqref{figB1} describe the possible heteroclinic sequences.  Again we see how the intermediate behaviour changes as the parameter $\mu$ changes.

\subsection{Inflation and Accelerated Expansion}
It follows that the deceleration parameter \eqref{def_q} can also be expressed in terms of the normalized bounded variables in this case as follows;
\begin{equation}
q= -\frac{1}{D^2} \left(-2 C \Sigma^2-2 \Psi^2 +\Phi^2-3\sqrt{\frac{3}{2}} \mu \sqrt{1-D^2} \Psi+\frac{3 \nu}{\sqrt{2}}  \Phi \Sigma  \right).
\end{equation}
The sign of the deceleration parameter indicates the nature of the expansionary evolution. If $q > 0$, then the cosmological expansion is decelerating, while negative values of $q$ indicate an accelerating or inflationary dynamics. See Table \eqref{Table2} for a summary of the sign of $q$ for each equilibrium point.

\subsection{Numerical Analysis}

\begin{figure}
$\begin{array}{rl}
\includegraphics[width=0.5\textwidth]{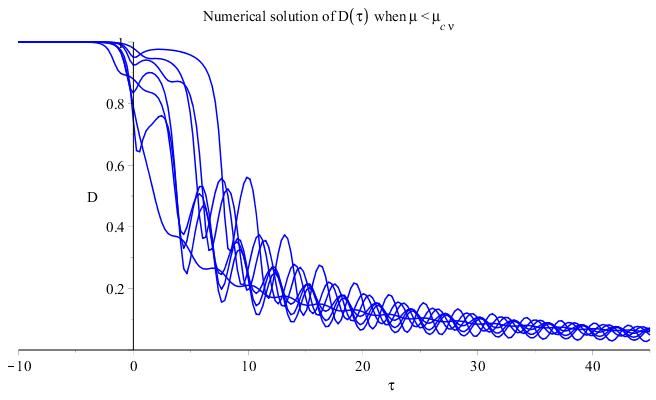} & \includegraphics[width=0.5\textwidth]{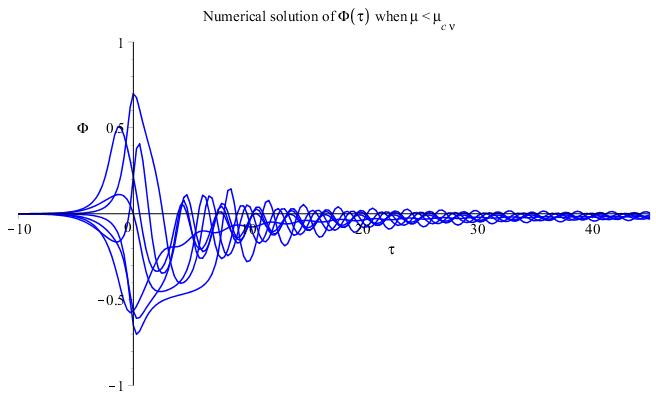} \\
\includegraphics[width=0.5\textwidth]{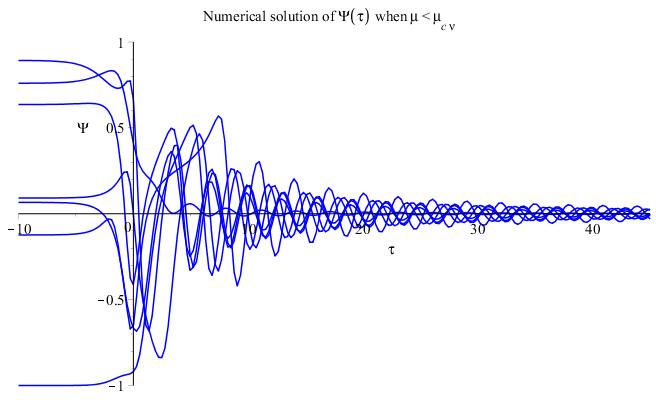}& \includegraphics[width=0.5\textwidth]{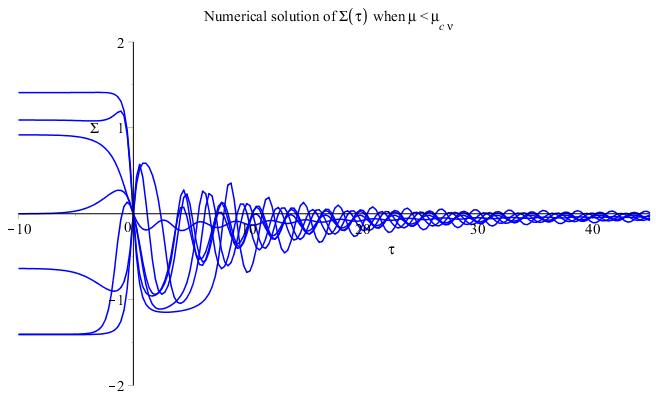}
\end{array}$
\caption{Numerical solutions of the system of differential equations \eqref{4De1}-\eqref{4De3} when $\mu<{\mu_{c\nu}}$. Note how the amplitude of the oscillations in all the variables slowly decreases. Note the future asymptotic behaviour is  $(D,\Phi,\Psi,\Sigma) \to (0, 0, 0, 0)$, $P_0$. There are a variety of past behaviour states which corresponds to the non-isolated set of equilibria given by $C^*$.  }\label{FIG-BV-MU-LESS}
\end{figure}

\begin{figure}
$\begin{array}{rl}
\includegraphics[width=0.5\textwidth]{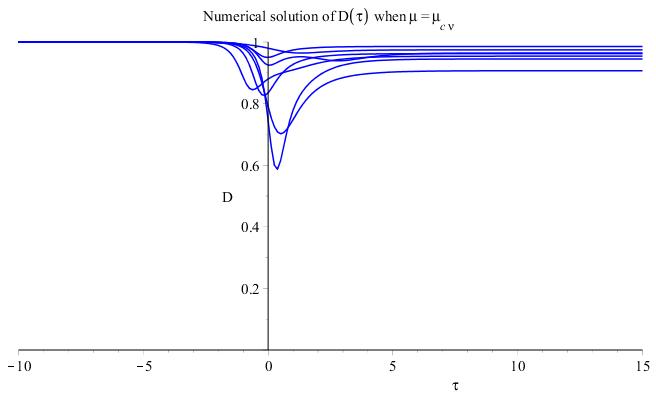} & \includegraphics[width=0.5\textwidth]{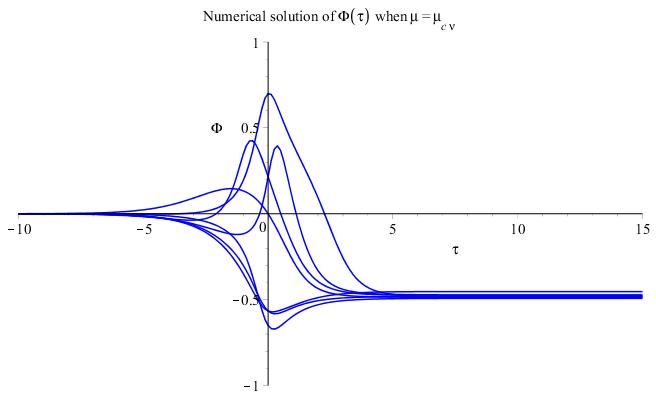} \\
\includegraphics[width=0.5\textwidth]{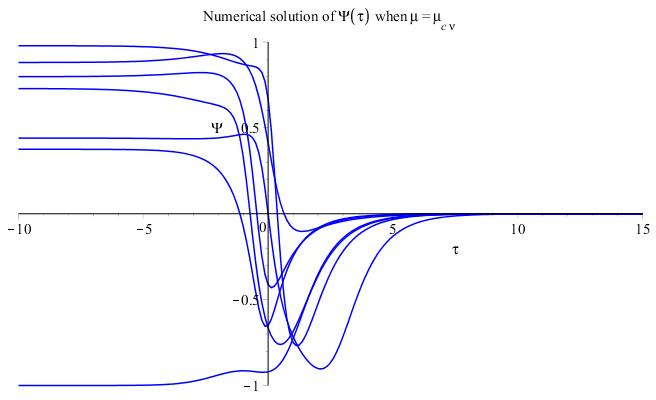}& \includegraphics[width=0.5\textwidth]{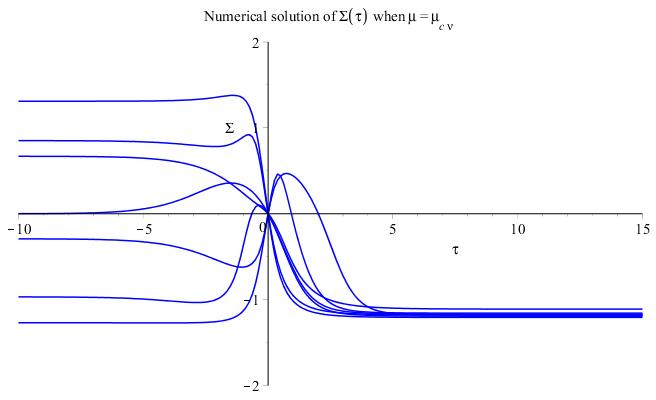}
\end{array}$
\caption{Numerical solutions of the system of differential equations \eqref{4De1}-\eqref{4De3} when $\mu={\mu_{c\nu}}$, the bifurcation value. Note the future asymptotic behaviour $\Psi\to 0$ but both $D$, $\Phi$, and $\Sigma$ approach different future asymptotic states depending on their initial condition. The future asymptotic state is $L_{04}$.  There are a variety of past behaviour states which corresponds to the non-isolated set of equilibria given by $C^*$.}\label{FIG-BV-MU-EQUAL}
\end{figure}

\begin{figure}
$\begin{array}{rl}
\includegraphics[width=0.5\textwidth]{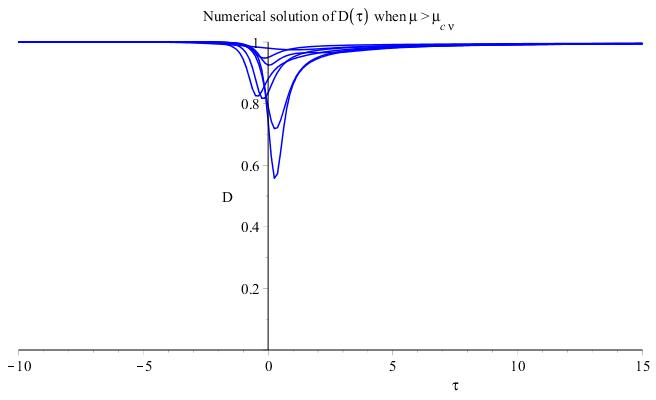} & \includegraphics[width=0.5\textwidth]{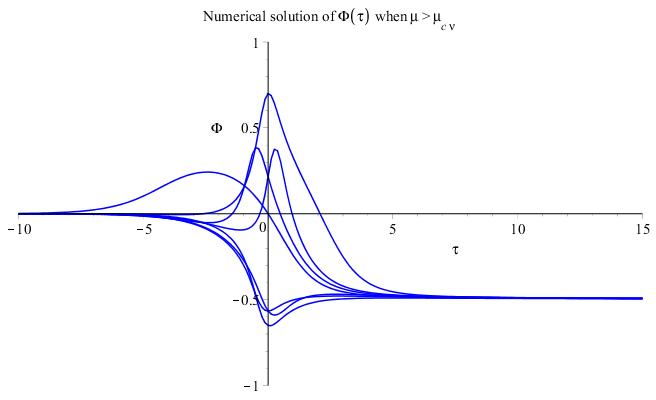} \\
\includegraphics[width=0.5\textwidth]{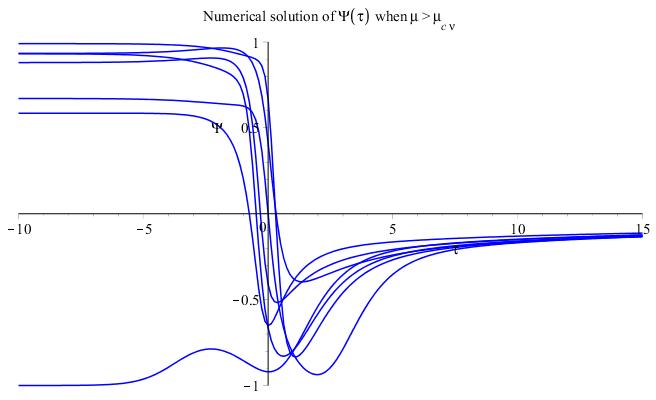}& \includegraphics[width=0.5\textwidth]{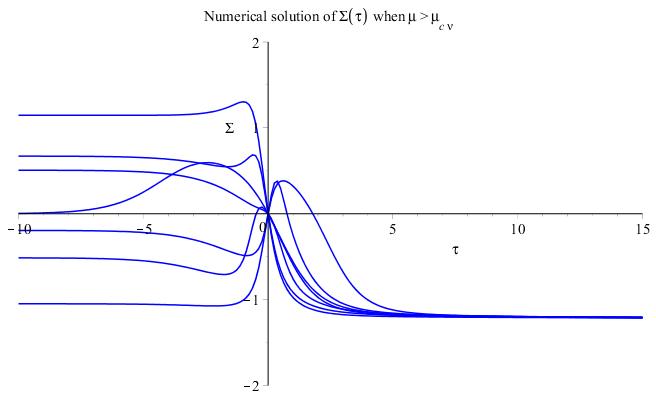}
\end{array}$
\caption{Numerical solutions of the system of differential equations \eqref{4De1}-\eqref{4De3} when $\mu>{\mu_{c\nu}}$. Note the future asymptotic behaviour is  $(D,\Phi,\Psi,\Sigma) \to P_4$ very slowly.  The timescale shown is for a short interval, but $\Psi$ in particular does indeed eventually converge to a value of $0$.  There are a variety of past behaviour states which corresponds to the non-isolated set of equilibria given by $C^*$. }\label{FIG-BV-MU-GREAT}
\end{figure}

It is constructive to illustrate a few numerical solutions for the three different regimes of future asymptotic behaviour, $\mu<{\mu_{c\nu}}$ [see Figure \eqref{FIG-BV-MU-LESS}], $\mu={\mu_{c\nu}}$ [see Figure \eqref{FIG-BV-MU-EQUAL}], and $\mu>{\mu_{c\nu}}$ [see Figure \eqref{FIG-BV-MU-GREAT}]. In each case the integrations are done in the full 4-dimensional phase space. The initial conditions are selected to show different past and future asymptotic behaviours and are the same as those used in the isotropic case, in that here we initially set $\Sigma(0)=0$.  We also do not show any phase portraits in this case as they are not as illustrative in higher dimensions as in the isotropic case.

\subsection{Observations}

In the spatially homogeneous and isotropic case analyzed here we observe that the past dynamics are independent of the strength of the coupling parameters $\mu$ and $\nu$.  We find in the zero curvature (Bianchi type I) and the negative curvature models (Bianchi type V) that the past asymptotic state is one which the anisotropy is non-trivial.  The past solution for both the zero and negative curvature models is the  Jacobs’ Bianchi type I non-vacuum massless scalar field solution \cite{Coley:2003mj}.

We also observe that the future asymptotic state depends on the strength of the coupling parameter $\mu$ and $\nu$.  For weak coupling of the Aether field to the scalar field, i.e., $ \mu < {\mu_{c \nu}}$, the dynamics are similar to but not the same as that found when there is no coupling of the Aether field to the scalar field, i.e., when $\mu=0$ and $\nu=0.$  Having $\nu>0$ drives the system towards intermediate states, $P_3$ and $P_4$, that are anisotropic in nature.  If $ \mu < {\mu_{c \nu}}$ then $P_0$ is the stable attractor: orbits oscillate and slowly decay in amplitude towards this final isotropic non-inflationary asymptotic state.

Similar to the isotropic case analyzed in Section \ref{section2}, for strong coupling of the Aether field to the scalar field, $ \mu > {\mu_{c \nu}}$, the dynamics are very different.  When $\mu>{\mu_{c\nu}}$ the scalar field does not come to rest at the minimum of the potential: the strength of the Aether interaction forces a different final equilibrium state. If $\mu > {\mu_{c \nu}}$ then we find that the stable equilibrium point in these models changes from the isotropic point $P_0$ to the anisotropic point $P_4$ if $\nu>0$ which is isotropic if $\nu=0$.  In this case, we find that the square of the scalar field, the square of the shear scalar, and the square of the expansion scalar scale together as
\begin{eqnarray*}
\frac{m^2}{2}\phi^2 &\sim& \frac{1}{3}\frac{\mu_2^2}{{\mu_{c\nu}}^2}(1+\theta^2)\\
\sigma^2 &\sim& \frac{1}{9}\frac{\nu^2}{C^2{\mu_{c\nu}}^2}(1+\theta^2)
\end{eqnarray*}
and consequently grow without bound.  We also observe that all orbits (excepting for the exceptional orbits) experience some period of accelerating expansion as they evolve to their final asymptotic state which is consistent with the isotropic case. The fundamental difference in the anisotropic case when $\mu > {\mu_{c \nu}}$ is that the future asymptotic state need not be isotropic.

Bianchi type V models in the standard inflationary scenario in GR $(C=1,\mu=0,\nu=0)$ isotropize as a rule. In GR, having interaction terms in the scalar field potential, equation \eqref{potential}, in which $(C=1, \mu>{\mu_{c\nu}},\nu>0)$, changes this rule to one in which the future asymptotic behaviour has accelerated expansion but is not isotropic. Similarly, in the Einstein Aether theory in the standard inflationary scenario $(0<C<1,\mu=0,\nu=0)$, one observes once again that the models will isotropize to the future.  However, just as in GR,  if $(0<C<1,\mu>{\mu_{c\nu}},\nu>0)$ then the models also have a future asymptotic behaviour which has accelerated expansion but is not isotropic.


\section{Discussion} \label{conclusion_1}

\subsection{Slow Roll Inflation}

In the standard slow roll inflationary scenario in GR, inflation occurs at intermediate times during a period of slow roll (in which $\ddot{\phi} \ll \theta \dot{\phi}$ and $\dot{\phi}^2 \ll \theta^2$) and where the anisotropy, if present, is insignificant when compared to the expansion.  However, the existence of a non-trivial coupling of the Aether field to the scalar field changes this scenario, and in particular if $\nu>0$, then there can be a significant departure from the standard scenario.

To find the slow roll inflationary attractor with scalar field/aether field coupling, we start off with all the same assumptions as above except one. The anisotropy is not assumed to be insignificant, but changes in the anisotropy are assumed to be small during slow roll inflation.  With these assumptions the slow roll solutions take on the form
\begin{eqnarray*}
\theta &=& \frac{3}{2}{\mu_{c\nu}}|\phi|\\
\sigma &=& \frac{\nu}{2c} \phi \\
\dot\phi &=& -{\mu_{c\nu}}\left(\mbox{sgn}(\phi)+\frac{\mu}{{\mu_{c\nu}}} \right)\\
\end{eqnarray*}
We see that if $\mu<{\mu_{c\nu}}$, then the slow roll solution is stable as $\phi$ and $\dot\phi$ have different signs (same for $\sigma$ and $\dot \sigma$). The number of e-foldings $N$ that can take place are:
\begin{eqnarray*}
N &=& \frac{1}{3}\int_{\phi_i}^{\phi_f}\,\frac{\theta}{\dot\phi} \,d\phi \\
  &=& \frac{1}{4(1+\mbox{sgn}(\phi)\frac{\mu}{\mu_{cv}})}\left(\phi_i^{\phantom{i}2}-\phi_f^{\phantom{i}2}\right)
\end{eqnarray*}
We note that if $\mbox{sgn}(\phi)<0$, then $N$ can be made large for fixed initial and final endpoints by choosing $\mu \lesssim {\mu_{c\nu}}$.
It appears that the existence of aether field/scalar field coupling terms of the nature studied here do not change the possibility of a period of slow-roll inflation at intermediate times.  We do note, however, that if $\nu>0$ then the inflation is anisotropic in nature.  Indeed, slow roll inflation is possible even when the mass of the scalar field is zero by simply choosing the coupling parameters to satisfy $\mu < {\mu_{c\nu}}|_{m=0}=\frac{\nu}{\sqrt{3C}}$.  In some sense, the slow roll inflationary expansion in this scenario is driven by the coupling to the shear.

\subsection{Final Comments}

We have investigated cosmological models in the Einstein-Aether theory in which scalar field matter is coupled to the aether through the scalar field potential. We have been especially interested in possible accelerated expansion and inflationary behaviour in a class of spatially homogeneous cosmological models. In particular, we have studied scalar field models in which the scalar-field potential depends on the time-like aether vector field through its expansion and shear.  We have observed that in the isotropic case, by choosing appropriate units, the dynamics are independent of the aether parameters.  The existence of the aether is to essentially re-normalize the gravitational constant $G_c=G/(1+c_\theta)$ in these cosmological settings. Further, in both the isotropic and anisotropic models studied here, we find that the past asymptotic state does not depend on the value of the aether parameters, $c_i$ or on the value of the scalar field coupling parameters $\mu$ and $\nu$.

However, in the anisotropic model the intermediate states and in some cases the final states do depend on a single combination of Aether parameters $c_\theta$ and $c_\sigma$ through the parameter $C$.  Indeed, these intermediate and final states increasingly become more anisotropic as a result of decreasing the Aether parameter $C$ from its maximum value $C=1$.  We also note that the future asymptotic states in both the isotropic and anisotropic models depend on the value of the scalar field/aether field coupling parameters $\mu$ and $\nu$ in the scalar field potential.  For sufficiently small values of the parameter $\mu$, the both the isotropic and anisotropic models experience a period of slow-roll inflation at intermediate times, even when the scalar field is massless. For sufficiently large values of the parameter $\mu$, the future asymptotic state changes to one which has accelerated expansion at late times.  Indeed, it is possible to have an accelerated expansion at late times even when the mass of the scalar field is zero, provided $\mu>0$ in the isotropic case, and $\mu>\nu/\sqrt{3C}$ in the anisotropic case.  In both of these isotropic and anisotropic models, the accelerated expansion at late times is a direct result of the scalar field/aether field coupling.  Further, it must be noted that in the anisotropic case, having a non-zero coupling parameter $\nu$ causes the future asymptotic state to be anisotropic.

The scalar field/aether field coupling parameters $\mu$ and $\nu$ in the scalar field potential modify the slow roll inflationary dynamics for a sufficiently small $\mu <{\mu_{c\nu}}$, which adds a driving force which can slow down or speed up (depending on the sign of scalar field initially) the slow roll inflation \cite{Solomon:2013iza,Donnelly:2010cr}.  In the anisotropic case there are further refinements to the slow roll regime (which can occur for a sufficiently small non-zero parameter $\nu$ in the potential).  Additionally, in the anisotropic case, the shear coupling causes the slow roll inflationary solution to be anisotropic in nature.

Recall, that a period of accelerated expansion is desirable at early intermediate times for inflationary purposes, but a period of accelerated expansion is also an attractive feature to have at late times to describe the effects of Dark Energy.  Here, in all cases (isotropic or anisotropic) or (zero curvature or negative curvature),  if $\mu$ is sufficiently small then there is a period of slow roll inflation at intermediate times, and if $\mu$ is sufficiently large, there will be accelerated expansion at late times.  Further, if $\nu>0$, these statements are true even when the scalar field is massless.


\acknowledgments This project is supported in part by the Atlantic Association for Research in the Mathematical Sciences through a Collaborative Research Grant.  We thank Theodore Kolokolnikov for his guidance on some technical points. BA would also like to thank the Government of Saudi Arabia for financial support. RvdH thanks the Department of Mathematics and Statistics at Dalhousie University for their kind hospitality.  AAC is supported by the Natural Sciences and Engineering Research Council of Canada.


\begin{thebibliography}{10}

\bibitem{Mattingly:2005re}
D.~Mattingly, \emph{{Modern tests of Lorentz invariance}},
  \href{http://dx.doi.org/10.12942/lrr-2005-5}{\emph{Living Rev. Rel.} {\bf 8}
  (2005) 5}, [\href{https://arxiv.org/abs/gr-qc/0502097}{{\tt gr-qc/0502097}}].

\bibitem{Jacobson:2000xp}
T.~Jacobson and D.~Mattingly, \emph{{Gravity with a dynamical preferred
  frame}}, \href{http://dx.doi.org/10.1103/PhysRevD.64.024028}{\emph{Phys.
  Rev.} {\bf D64} (2001) 024028},
  [\href{https://arxiv.org/abs/gr-qc/0007031}{{\tt gr-qc/0007031}}].

\bibitem{Liberati:2013xla}
S.~Liberati, \emph{{Tests of Lorentz invariance: a 2013 update}},
  \href{http://dx.doi.org/10.1088/0264-9381/30/13/133001}{\emph{Class. Quant.
  Grav.} {\bf 30} (2013) 133001}, [\href{https://arxiv.org/abs/1304.5795}{{\tt
  1304.5795}}].

\bibitem{Jacobson:2008aj}
T.~Jacobson, \emph{{Einstein-aether gravity: A Status report}}, {\emph{PoS}
  {\bf QG-PH} (2007) 020}, [\href{https://arxiv.org/abs/0801.1547}{{\tt
  0801.1547}}].

\bibitem{Zlosnik:2006zu}
T.~G. Zlosnik, P.~G. Ferreira and G.~D. Starkman, \emph{{Modifying gravity with
  the Aether: An alternative to Dark Matter}},
  \href{http://dx.doi.org/10.1103/PhysRevD.75.044017}{\emph{Phys. Rev.} {\bf
  D75} (2007) 044017}, [\href{https://arxiv.org/abs/astro-ph/0607411}{{\tt
  astro-ph/0607411}}].

\bibitem{Clifton:2011jh}
T.~Clifton, P.~G. Ferreira, A.~Padilla and C.~Skordis, \emph{{Modified Gravity
  and Cosmology}},
  \href{http://dx.doi.org/10.1016/j.physrep.2012.01.001}{\emph{Phys. Rept.}
  {\bf 513} (2012) 1--189}, [\href{https://arxiv.org/abs/1106.2476}{{\tt
  1106.2476}}].

\bibitem{Jacobson:2004ts}
T.~Jacobson and D.~Mattingly, \emph{{Einstein-Aether waves}},
  \href{http://dx.doi.org/10.1103/PhysRevD.70.024003}{\emph{Phys. Rev.} {\bf
  D70} (2004) 024003}, [\href{https://arxiv.org/abs/gr-qc/0402005}{{\tt
  gr-qc/0402005}}].

\bibitem{Jacobson:2010mxa}
T.~Jacobson, \emph{{Extended Horava gravity and Einstein-aether theory}},
  \href{http://dx.doi.org/10.1103/PhysRevD.82.129901,
  10.1103/PhysRevD.81.101502}{\emph{Phys. Rev.} {\bf D81} (2010) 101502},
  [\href{https://arxiv.org/abs/1001.4823}{{\tt 1001.4823}}].

\bibitem{Jacobson:2010mxb}
T.~Jacobson, \emph{Erratum: Extended ho\ifmmode \check{r}\else \v{r}\fi{}ava
  gravity and einstein-aether theory [phys. rev. d 81, 101502 (2010)]},
  \href{http://dx.doi.org/10.1103/PhysRevD.82.129901}{\emph{Phys. Rev. D} {\bf
  82} (Dec, 2010) 129901}.

\bibitem{Garfinkle:2007bk}
D.~Garfinkle, C.~Eling and T.~Jacobson, \emph{{Numerical simulations of
  gravitational collapse in Einstein-aether theory}},
  \href{http://dx.doi.org/10.1103/PhysRevD.76.024003}{\emph{Phys. Rev.} {\bf
  D76} (2007) 024003}, [\href{https://arxiv.org/abs/gr-qc/0703093}{{\tt
  gr-qc/0703093}}].

\bibitem{Garfinkle:2011iw}
D.~Garfinkle and T.~Jacobson, \emph{{A positive energy theorem for
  Einstein-aether and Ho\v{r}ava gravity}},
  \href{http://dx.doi.org/10.1103/PhysRevLett.107.191102}{\emph{Phys. Rev.
  Lett.} {\bf 107} (2011) 191102}, [\href{https://arxiv.org/abs/1108.1835}{{\tt
  1108.1835}}].

\bibitem{Barrow:2012qy}
J.~D. Barrow, \emph{{Some Inflationary Einstein-Aether Cosmologies}},
  \href{http://dx.doi.org/10.1103/PhysRevD.85.047503}{\emph{Phys. Rev.} {\bf
  D85} (2012) 047503}, [\href{https://arxiv.org/abs/1201.2882}{{\tt
  1201.2882}}].

\bibitem{Sandin:2012gq}
P.~Sandin, B.~Alhulaimi and A.~Coley, \emph{{Stability of Einstein-Aether
  Cosmological Models}},
  \href{http://dx.doi.org/10.1103/PhysRevD.87.044031}{\emph{Phys. Rev.} {\bf
  D87} (2013) 044031}, [\href{https://arxiv.org/abs/1211.4402}{{\tt
  1211.4402}}].

\bibitem{Donnelly:2010cr}
W.~Donnelly and T.~Jacobson, \emph{{Coupling the inflaton to an expanding
  aether}}, \href{http://dx.doi.org/10.1103/PhysRevD.82.064032}{\emph{Phys.
  Rev.} {\bf D82} (2010) 064032}, [\href{https://arxiv.org/abs/1007.2594}{{\tt
  1007.2594}}].

\bibitem{Solomon:2013iza}
A.~R. Solomon and J.~D. Barrow, \emph{{Inflationary Instabilities of
  Einstein-Aether Cosmology}},
  \href{http://dx.doi.org/10.1103/PhysRevD.89.024001}{\emph{Phys. Rev.} {\bf
  D89} (2014) 024001}, [\href{https://arxiv.org/abs/1309.4778}{{\tt
  1309.4778}}].

\bibitem{Solomon:2015hja}
A.~R. Solomon, \emph{{Cosmology Beyond Einstein}}.
\newblock PhD thesis, Cambridge U., Cham, 2015.
\newblock \href{https://arxiv.org/abs/1508.06859}{{\tt 1508.06859}}.
\newblock 10.1007/978-3-319-46621-7.

\bibitem{Alhulaimi:2017}
B.~Alhulaimi, \emph{Einstein-aether Cosmological Scalar Field Models}.
\newblock PhD thesis, Dalhousie University, Halifax, Nova Scotia, February,
  2017.

\bibitem{Kanno:2006ty}
S.~Kanno and J.~Soda, \emph{{Lorentz Violating Inflation}},
  \href{http://dx.doi.org/10.1103/PhysRevD.74.063505}{\emph{Phys. Rev.} {\bf
  D74} (2006) 063505}, [\href{https://arxiv.org/abs/hep-th/0604192}{{\tt
  hep-th/0604192}}].

\bibitem{Olive:1989nu}
K.~A. Olive, \emph{{Inflation}},
  \href{http://dx.doi.org/10.1016/0370-1573(90)90144-Q}{\emph{Phys. Rept.} {\bf
  190} (1990) 307--403}.

\bibitem{Linde:1987}
A.~D. Linde, \emph{{Inflation and Quantum Cosmology}},  in \emph{{300 Years of
  Gravity}}, pp.~604--630, 1987.

\bibitem{Riess:1998cb}
{\scshape Supernova Search Team} collaboration, A.~G. Riess et~al.,
  \emph{{Observational evidence from supernovae for an accelerating universe
  and a cosmological constant}},
  \href{http://dx.doi.org/10.1086/300499}{\emph{Astron. J.} {\bf 116} (1998)
  1009--1038}, [\href{https://arxiv.org/abs/astro-ph/9805201}{{\tt
  astro-ph/9805201}}].

\bibitem{Perlmutter:1998np}
{\scshape Supernova Cosmology Project} collaboration, S.~Perlmutter et~al.,
  \emph{{Measurements of Omega and Lambda from 42 high redshift supernovae}},
  \href{http://dx.doi.org/10.1086/307221}{\emph{Astrophys. J.} {\bf 517} (1999)
  565--586}, [\href{https://arxiv.org/abs/astro-ph/9812133}{{\tt
  astro-ph/9812133}}].

\bibitem{Coley:2015qqa}
A.~A. Coley, G.~Leon, P.~Sandin and J.~Latta, \emph{{Spherically symmetric
  Einstein-aether perfect fluid models}},
  \href{http://dx.doi.org/10.1088/1475-7516/2015/12/010}{\emph{JCAP} {\bf 12}
  (2015) 010}, [\href{https://arxiv.org/abs/1508.00276}{{\tt 1508.00276}}].

\bibitem{Latta:2016jix}
J.~Latta, G.~Leon and A.~Paliathanasis, \emph{{Kantowski-Sachs Einstein-æther
  perfect fluid models}},
  \href{http://dx.doi.org/10.1088/1475-7516/2016/11/051}{\emph{JCAP} {\bf 1611}
  (2016) 051}, [\href{https://arxiv.org/abs/1606.08586}{{\tt 1606.08586}}].

\bibitem{Coley:2003mj}
A.~A. Coley, \emph{{Dynamical systems and cosmology}}, vol.~291.
\newblock Kluwer, Dordrecht, Netherlands, 2003,
  \href{http://dx.doi.org/10.1007/978-94-017-0327-7}{10.1007/978-94-017-0327-7}.

\bibitem{wainwright_ellis2005}
J.~Wainwright and G.~Ellis, \emph{Dynamical Systems in Cosmology}.
\newblock Cambridge University Press, 2005.

\bibitem{hinch1991}
E.~Hinch, \emph{Perturbation Methods}.
\newblock Cambridge Texts in Applied Mathematics. Cambridge University Press,
  1991.

\bibitem{kevorkian2013}
J.~Kevorkian and J.~Cole, \emph{Perturbation Methods in Applied Mathematics}.
\newblock Applied Mathematical Sciences. Springer New York, 2013.

\bibitem{nayfeh2000}
A.~Nayfeh, \emph{Perturbation methods}.
\newblock Physics textbook. John Wiley \& Sons, 2000.

\bibitem{wiggins2003}
S.~Wiggins, \emph{{Introduction to Applied Nonlinear Dynamical Systems and
  Chaos}}.
\newblock Texts in Applied Mathematics. Springer New York, 2003.

\bibitem{Alhulaimi:2013sha}
B.~Alhulaimi, A.~Coley and P.~Sandin, \emph{{Anisotropic Einstein-aether
  cosmological models}}, \href{http://dx.doi.org/10.1063/1.4802246}{\emph{J.
  Math. Phys.} {\bf 54} (2013) 042503}.

\bibitem{Carruthers:2010ii}
I.~Carruthers and T.~Jacobson, \emph{{Cosmic alignment of the aether}},
  \href{http://dx.doi.org/10.1103/PhysRevD.83.024034}{\emph{Phys. Rev.} {\bf
  D83} (2011) 024034}, [\href{https://arxiv.org/abs/1011.6466}{{\tt
  1011.6466}}].

\bibitem{Foster:2005dk}
B.~Z. Foster and T.~Jacobson, \emph{{Post-Newtonian parameters and constraints
  on Einstein-aether theory}},
  \href{http://dx.doi.org/10.1103/PhysRevD.73.064015}{\emph{Phys. Rev.} {\bf
  D73} (2006) 064015}, [\href{https://arxiv.org/abs/gr-qc/0509083}{{\tt
  gr-qc/0509083}}].

\end{thebibliography}
\providecommand{\href}[2]{#2}\begingroup\raggedright\endgroup


\appendix
\section{Appendix:  Constraints on the Einstein Aether Parameters $c_i$} \label{Appendix}

According to \cite{Foster:2005dk,Jacobson:2008aj} the PPN parameters for Einstein's General Relativity and Einstein Aether theory are identical if
\begin{eqnarray}
c_2 &=& \frac{-2c_1^2-c_1c_3+c_3^2}{3c_1}, \nonumber\\
c_4 &=& -\frac{c_3^2}{c_1}.\label{PPN}
\end{eqnarray}
If one also assumes the squared speeds of massless modes relative to the Aether rest frame must be super-luminal to avoid vacuum \v{C}erenkov radiation \cite{Foster:2005dk,Jacobson:2008aj}, then two additional constraints must be satisfied
\begin{eqnarray}
0 &\leq& c_1+c_3 \leq \frac{1}{2}, \nonumber\\
0 &\leq& c_1-c_3 \leq \frac{c_1+c_3}{3[1-2(c_1+c_3)]},\label{Cerenkov}
\end{eqnarray}
which when combined with \eqref{PPN} are sufficient to show both positive energy modes and linear stability.  Recall that the $c_i$ employed here are one-half of the values used in \cite{Foster:2005dk,Jacobson:2008aj}. In terms of $c_\sigma$ and $c_\theta$ used in this paper, assuming that Einstein Aether and GR are equivalent for weak fields, i.e., equation \eqref{PPN}, then these constraints \eqref{Cerenkov} become
\begin{eqnarray*}
0 &\leq& c_\sigma \leq \frac{1}{2} \\
\frac{c_\sigma}{3c_\sigma-2} &\leq& c_\theta \leq 0
\end{eqnarray*}
and in terms of $c_\sigma$ and $C$
\begin{eqnarray}
0 &\leq& c_\sigma \leq \frac{1}{2}, \nonumber \\
1-2c_\sigma &\leq& C \leq 1-\frac{3}{2}c_\sigma. \label{C_constraints}
\end{eqnarray}

\begin{figure}
\begin{center}
\includegraphics[scale=0.4]{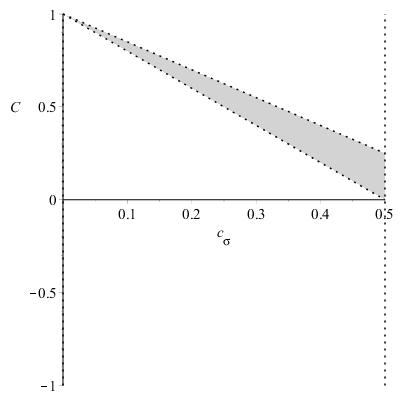}
\caption{Range of allowable parameter values for the aether parameters $c_\sigma$ and $C=\frac{1-2c_\sigma}{1+c_\theta}$. Recall, to ensure a compact phase space we assumed that $C>0$, which fortunately agrees with the set of allowable parameter values via a PPN analysis.  GR is represented by the point $(c_\sigma,C)=(0,1)$.}
\end{center}
\end{figure}

Donnelly and Jacobson included the coupling parameter $\mu$ in their analysis of the parameter constraints and showed that these constraints are relaxed when $\mu>0$ \cite{Donnelly:2010cr} and are automatically satisfied when the PPN parameters match those of GR.

\end{document}